\documentclass[preprint,aps,nofootinbib]{revtex4}
\input{epsf.tex}
\usepackage[dvips]{graphics,epsfig}
\usepackage{textcomp}
\def\bkR{{\rm I\kern-.17em R}}
\def\bkC{{\rm \kern.24em \vrule width.05em height1.4ex depth-.05ex \kern-.26em C}}
\def\to{\rightarrow}

\def\be{\beta}

\def\half{{\textstyle{1\over 2}}}
\def\frac#1#2{{\textstyle{{#1}\over {#2}}}}

\def\lsim{\mathrel{\rlap{\lower4pt\hbox{\hskip1pt$\sim$}}
    \raise1pt\hbox{$<$}}}
\def\gsim{\mathrel{\rlap{\lower4pt\hbox{\hskip1pt$\sim$}}
    \raise1pt\hbox{$>$}}}
\def\sqr#1#2{{\vcenter{\vbox{\hrule height.#2pt
         \hbox{\vrule width.#2pt height#1pt \kern#1pt
         \vrule width.#2pt}
         \hrule height.#2pt}}}}

\def\laq{\raise 0.4 ex \hbox{$<$}\kern -0.8 em\lower 0.62 ex\hbox{$\sim$}}
\def\gaq{\raise 0.4 ex \hbox{$>$}\kern -0.7 em\lower 0.62 ex\hbox{$\sim$}}

\def\be{\begin{equation}}
\def\ee{\end{equation}}
\def\ba{\begin{eqnarray}}
\def\ea{\end{eqnarray}}
\def\Box{\mathord{\dalemb{7.9}{8}\hbox{\hskip1pt}}}
\def\dalemb#1#2{{\vbox{\hrule height.#2pt
        \hbox{\vrule width.#2pt height#1pt \kern#1pt \vrule width.#2pt}
        \hrule height.#2pt}}}

\def\Box{\mathord{\dalemb{7.9}{8}\hbox{\hskip1pt}}}
\def\dalemb#1#2{{\vbox{\hrule height.#2pt
        \hbox{\vrule width.#2pt height#1pt \kern#1pt \vrule width.#2pt}
        \hrule height.#2pt}}}

\def\ba{\begin{eqnarray}}
\def\ea{\end{eqnarray}}
\def\be{\begin{equation}}
\def\ee{\end{equation}}

\def\gtorder{\mathrel{\raise.3ex\hbox{$>$}\mkern-14mu
             \lower0.6ex\hbox{$\sim$}}}
\def\ltorder{\mathrel{\raise.3ex\hbox{$<$}\mkern-14mu
             \lower0.6ex\hbox{$\sim$}}}

\begin{document}

\rightline{DM/IST-6.2006}
\rightline{November 2006}

\title{Weyl-Wigner Formulation of Noncommutative Quantum Mechanics}

\author{Catarina Bastos\footnote{Also at Centro de F\'\i sica dos Plasmas,
IST. Email address: cbastos@fisica.ist.utl.pt}, Orfeu Bertolami\footnote{Also at Centro de F\'\i sica dos Plasmas,
IST. Email address: orfeu@cosmos.ist.utl.pt}}

\vskip 0.3cm

\affiliation{Departamento de F\'\i sica, Instituto Superior T\'ecnico \\
Avenida Rovisco Pais 1, 1049-001 Lisboa, Portugal}

\author{Nuno Costa Dias\footnote{Also at Grupo de F\'{\i}sica Matem\'atica, UL,
Av. Prof. Gama Pinto 2, 1649-003, Lisboa, Portugal.
Email address: ncdias@mail.telepac.pt}, Jo\~ao Nuno Prata\footnote{Also at Grupo de F\'{\i}sica Matem\'atica, UL,
Av. Prof. Gama Pinto 2, 1649-003, Lisboa, Portugal.
Email address:joao.prata@mail.telepac.pt}}

\vskip 0.3cm

\affiliation{Departamento de Matem\'{a}tica, Universidade Lus\'ofona de
Humanidades e Tecnologias \\
Av. Campo Grande, 376, 1749-024 Lisboa, Portugal}


\vskip 0.5cm

\begin{abstract}

\vskip 1cm

{We address the phase space formulation of a noncommutative extension of quantum mechanics in arbitrary dimension, displaying both spatial and momentum noncommutativity. By resorting to a covariant generalization of the Weyl-Wigner transform and to the Seiberg-Witten map we
construct an isomorphism between the operator and the phase space representations of the extended Heisenberg algebra. This map provides a systematic approach to derive the entire structure of noncommutative quantum mechanics in phase space. We construct the extended
starproduct, Moyal bracket and propose a general definition of noncommutative states. We study the dynamical and eigenvalue equations of the theory and prove that the entire formalism is independent of the particular choice of Seiberg-Witten map. Our approach unifies and
generalizes all the previous proposals for the phase space formulation of noncommutative quantum mechanics. For concreteness we rederive these proposals by restricting our formalism to some 2-dimensional spaces.}

\end{abstract}

\maketitle

\section{Introduction}

Noncommutative extensions of quantum mechanics have been recently widely discussed in the literature \cite{Ho,Nair,Duval,Horvathy,Zhang_1,Zhang_2,Demetrian,Gamboa,Li,Acatrinei,Bertolami1,Bertolami2,Dias5,Bastos}. This interest has its roots in the noncommutative field theories and their connection with quantum gravity and string theory. It is widely believed that the final theory of quantum gravity will determine the fundamental structure of space-time and does contain some sort of noncommutative structure. This will have of course, a profound impact on the mathematical foundation of quantum mechanics and quantum field theory and will certainly have implications on their predictions. The quest to find deviations from the predictions of standard quantum mechanics, that could be regarded as a signature of quantum gravity, is one of the key motivations for the recent interest in noncommutative quantum mechanics (NCQM), the nonrelativistic one-particle sector of noncommutative field theories.

Most of the models of NCQM considered in the literature are based on canonical extensions of the  Heisenberg algebra. In these models time is
required to be a commutative parameter. In a $d$ dimensional space with noncommuting position and momentum variables, the extended Heisenberg
algebra reads:
\begin{equation}
\left[\hat q_i, \hat q_j \right] = i \theta_{ij} , \hspace{0.5 cm} \left[\hat q_i, \hat p_j \right] = i \hbar \delta_{ij} ,
\hspace{0.5 cm} \left[\hat p_i, \hat p_j \right] = i \eta_{ij} ,  \hspace{0.5 cm} i,j= 1, ... ,d
\label{Eq1.1}
\end{equation}
where $\eta_{ij}$ and $\theta_{ij}$ are antisymmetric real constant ($d \times d$) matrices and $\delta_{ij}$ is the identity matrix.
Theoretical predictions for specific noncommutative systems have been compared with experimental data leading to bounds on the noncommutative parameters \cite{Bertolami1,Carroll}:
\begin{equation}
\theta \leq 4 \times 10^{-40} m^2 \hspace{0.5 cm},\hspace{0.5 cm} \eta \leq 1.76 \times 10^{-61} kg^2 m^2 s^{-2}.
\label{Eq1.2}
\end{equation}
Moreover, a great deal of work has been devoted to structural and formal aspects of the quantum theory based on the algebra (\ref{Eq1.1}).
The key property of the extended Heisenberg algebra is that it is related to the standard Heisenberg algebra:
\begin{equation}
\left[\hat R_i, \hat R_j \right] = 0 , \hspace{0.5 cm} \left[\hat R_i, \hat \Pi_j \right]
= i \hbar \delta_{ij} , \hspace{0.5 cm} \left[\hat \Pi_i, \hat \Pi_j \right] = 0 ,
\hspace{0.5 cm} i,j= 1, ... ,d ~,
\label{Eq1.3}
\end{equation}
by a class of linear (non-canonical) transformations:
\begin{equation}
\hat q_i = \hat q_i \left(\hat R_j , \hat \Pi_j \right) \hspace{1 cm}
\hat p_i = \hat p_i \left(\hat R_j , \hat \Pi_j \right)
\label{Eq1.4}
\end{equation}
which are often refered to as the Seiberg-Witten (SW) maps \cite{Seiberg}. By resorting to one of these transformations, one is able to find a representation of the noncommutative observables as operators acting on the conventional Hilbert space of ordinary quantum mechanics, i.e. to convert the noncommutative system into a modified commutative system which exhibits an explicit dependence on the noncommutative parameters as well as on the particular SW map. The states of the system are then wave functions of the ordinary Hilbert space and its dynamics is determined by the Schr\"{o}dinger equation with a modified $\eta,\theta$-dependent Hamiltonian. Although the entire formulation is dependent of the particular SW map used to perform the noncommutative-commutative conversion, this is not so for the physical predictions such as expectation values and probability distributions. Still, the fact that the formalism is not manifestly invariant under a modification of the SW map is a disadvantage. Moreover, both the observables and the states, being written in terms of the Heisenberg variables (\ref{Eq1.3}), do not  display a simple mathematical structure. This tends to obscure their physical meaning.

The deformation quantization method \cite{Groenewold,Moyal,Baker,Wigner,Bayen,Dias1,Dias2,Vey,Dias3,Dias4,Segal,Pool,Dubin,Wilde,Bordemann,Fedosov1,Fedosov2,Kontsevich} leads to a phase space formulation of quantum mechanics alternative to the more conventional path integral and operator formulations. This is a powerful quantization procedure that can be applied to classical systems defined on arbitrary non-flat Poisson or sympletic manifolds. In the flat case it leads to the well known Weyl-Wigner formulation of quantum mechanics which is tantamount to a phase space representation of the Heisenberg algebra (\ref{Eq1.3}). This formulation of quantum mechanics is akin to classical statistical mechanics. Observables are represented by phase space functions and states by a quasi-distribution known as the Wigner function. The key algebraic structure of the theory is an associative and non-commutative $\star$-product which carries the information about the commutation relations (\ref{Eq1.3}).

The most natural representation of the algebra (\ref{Eq1.1}) is also in phase space where the commutation relations (\ref{Eq1.1}) can be implemented in a Lie algebraic way, i.e. following from a noncommutative and associative extended $\star$-product of phase space functions: $i \hbar [A,B]=A\star B-B\star A$. This formulation leads directly to a noncommutative extension of the Weyl-Wigner formulation of quantum mechanics. Some work concerning the deformation quantization of NCQM has recently appeared in the literature. Various authors proposed formulae for the quasi-distribution on the plane with only spatial noncommutativity \cite{Jing,Rosenbaum1,Muthukumar} or with phase space noncommutativity \cite{Rosenbaum2}.

The former approaches point in an interesting direction. However, they have been developed for a particular dimension $(d=2)$ and for specific values of the noncommutative parameters and it is difficult to see how they relate to each other and to the operator representations of the noncommutative algebra. A global formalism from where these formulations could be derived as particular cases is still missing. One of the main issues is the absence of a unified point of view on the definition of state for phase space NCQM. A clear relation with the operator formulation of NCQM is also quite relevant and is still lacking and so is a definitive proof that the formalism is independent of the particular choice of the SW map.

In this paper we will study the phase space formulation of NCQM and address these issues. Our main tool will be the covariant generalization of the Weyl-Wigner map \cite{Dias3}. In its original formulation the non-covariant Weyl-Wigner map is an isomorphism between the operator and the phase space representations of the ordinary quantum mechanics based on the algebra (\ref{Eq1.3}) and it provides the simplest approach to derive most of the mathematical structure of conventional phase space quantum mechanics \cite{Groenewold,Moyal,Baker,Wigner,Bayen,Dias1,Dias2,Dias4}. The covariant extension of this map was studied in \cite{Dias3} in connection with a diffeomorphism invariant formulation of Weyl-Wigner quantum mechanics \cite{Vey}.

By resorting to the covariant generalization of the Weyl-Wigner transform and to the SW map we will construct an extended Weyl-Wigner map for noncommutative quantum mechanics. This map is an isomorphism between the operator and phase space representations of the extended Heisenberg algebra (\ref{Eq1.1}) and, therefore, it provides a systematic approach to derive the phase space formulation of NCQM in its most general form, i.e. for arbitrary $d$-dimensions and any value of the noncommutative parameters.
By applying it to the density matrix, we get the noncommutative counterpart of the Wigner function, which we shall call the noncommutative Wigner function. It represents the state of the system. Moreover, this extended Weyl-Wigner transform can be applied to derive the fundamental algebraic structure of the theory. We shall use it to: 

(i) Get the extended $\star$-product and a Moyal bracket encompassing the noncommutativity of the theory and 

(ii) Derive the dynamical and eigenvalue equations for NCQM. 

Finally, we will show that the noncommutative Weyl-Wigner transform, and therefore the entire formulation of NCQM, does not depend on the particular choice for the SW map.

Our approach provides a unified framework for the phase space formulation of NCQM. The entire structure of the theory follows directly from the extended Weyl-Wigner map. Previous approaches to phase space NCQM are generalized and unified. We shall re-derive some of these formulations by restricting our formalism to their particular physical situations.

This paper is organized as follows. In section 2, we discuss the general features of the Weyl-Wigner map and its covariant generalization. We also make a brief review of the deformation quantization method. In section 3, we construct the extended star-product, the extended Moyal bracket and the noncommutative Wigner function, and extensively discuss their properties. We also prove that the extended Weyl-Wigner map is independent of the particular choice of the SW transformation. In section 4, we specialise our findings to 2 dimensions and compare them with results already known in the literature. Section 5, summarizes our conclusions.

\section{The generalized Weyl-Wigner map}

In this section we aim to review the main structures of the covariant formulation of deformation quantization
with emphasis on the generalized Weyl-Wigner transform. Our analysis is restricted to the case of flat phase spaces.
The formalism presented here will help to make the forthcoming sections
simpler and more self-contained.
The reader is referred to Refs. \cite{Bayen,Dias1,Dias2,Vey,Dias3,Dias4,Segal,Pool,Dubin} for a more detailed presentation
and to Refs. \cite{Wilde,Bordemann,Fedosov1,Fedosov2,Kontsevich} for the generalization of the formalism to the non-flat case.

The general interest of the covariant version of the deformation quantization procedure is that it leads
to a larger set of quantum phase space representations. We shall see in the next section that from
an operational point of view noncommutative quantum mechanics in phase space is a covariant phase
space quantum mechanics with a particular choice of non-canonical coordinates.

Let us then settle down the preliminaries: we consider a $d$-dimensional dynamical system, such that its
classical formulation lives in the flat phase
space $T^*M \simeq \bkR^{2d}$. A set of global canonical coordinates
\begin{eqnarray}
\xi&=&(R,\Pi)=(R_i, \Pi_i),\,\, i=1, \cdots,d,\nonumber\\
\xi_{\alpha}=R_{\alpha},\,\, \alpha&=&1,..,d \hspace{0.2cm} \mbox{and} \quad \xi_{\alpha}=\Pi_{\alpha-d}, \,\,\alpha=d+1,..,2d
\label{Eq2.1}
\end{eqnarray}
can then be defined in $T^*M$ in terms of which the sympletic structure reads $w=dR_i \wedge d\Pi_i$.
In the sequel the Latin letters run from $1$ to $d$ (e.g. $i,j,k, \cdots = 1, \cdots, d$), whereas the
Greek letters stand for phase space indices (e.g. $\alpha, \beta, \gamma, \cdots = 1, \cdots, 2d$), unless
otherwise stated. Moreover, summation over repeated indices is assumed.
Upon quantization, the set $\{\hat{\xi}_{\alpha},\, \alpha = 1,..,2d\}$ satisfies the commutation relations of
the standard Heisenberg algebra and $\{\hat R_i,\, i=1,..,d \}$ constitutes a
complete set of commuting observables. Let us denote by $|R\rangle$ the general eigenstate of $\hat R$ associated
to the array of eigenvalues $R_i,\, i=1,..,d$ and spanning the Hilbert space ${\cal H}=L_2(\bkR^d,dR)$ of
complex valued functions $\psi :\bkR^{d}\longrightarrow \bkC$ $(\psi (R) = \langle R| \psi\rangle)$, which are square integrable with respect to
the standard Lesbegue measure $dR$.
The scalar product in ${\cal H}$ is given by:
\begin{equation}
(\psi,\phi)_{\cal H}=\int dR \,\overline{\psi(R)} \phi(R)
\label{Eq2.2}
\end{equation}
where the over-bar denotes complex conjugation.\\

\noindent We now introduce the Gel'fand triple of vector spaces
\cite{Gelfand,Roberts,Antoine,Bohm}:
\begin{equation}
{\cal S}(\bkR^d) \subset {\cal H} \subset {\cal S}'(\bkR^d),
\label{Eq2.3}
\end{equation}
where ${\cal S}(\bkR^d)$ is the space of all complex valued functions $t(R)$ that are infinitely smooth and, as $||R|| \to \infty$, they and all their partial derivatives decay to zero faster than any power of $1/||R||$. ${\cal S}(\bkR^d)$ is the space of Schwartz test functions or of rapid descent test functions \cite{Hormander,Zemanian}, and ${\cal S}'(\bkR^d)$ is its dual, i.e. the space of tempered distributions. In analogy with (\ref{Eq2.3}) let us also introduce the triple:
\begin{equation}
{\cal S}(\bkR^{2d}) \subset {\cal F}=L_2(\bkR^{2d},dRd\Pi) \subset {\cal S}'(\bkR^{2d}),
\label{Eq2.4}
\end{equation}
where ${\cal F} $ is the set of square integrable phase space functions with scalar product:
\begin{equation}
(F,G)_{\cal F}={1\over(2\pi \hbar)^d} \int dR \int d\Pi \,\, \overline{F(R,\Pi)} G(R,\Pi).
\label{Eq2.5}
\end{equation}
Finally, let $\hat{\cal S}'$ be the set of linear operators admitting a representation of the form \cite{Dubin}:
\begin{equation}
\hat A: {\cal S}(\bkR^d) \longrightarrow {\cal S}'(\bkR^d); \, \, \psi(R) \longrightarrow (\hat A \psi)(R)= \int dR' \, A_K(R,R') \psi(R'),
\label{Eq2.6}
\end{equation}
where $A_K(R,R')=\langle R|\hat A|R'\rangle \in {\cal S}'(\bkR^{2d})$ is a distributional kernel.
The elements of $\hat{\cal S}'$ are named generalized operators.\\

\noindent
The Weyl-Wigner transform is the one-to-one invertible map \cite{Segal,Dubin,Bracken}:
\begin{eqnarray}
W_{\xi}: \hat{\cal S}'  & \longrightarrow & {\cal S}'(\bkR^{2d});  \nonumber\\
\hat A & \longrightarrow & A(R,\Pi)= W_{\xi}(\hat A) = \hbar^d  \int dy \, e^{-i\Pi\cdot y}
A_K(R+\frac{\hbar}{2} y,R-\frac{\hbar}{2} y) \nonumber \\
&&\hspace{3.2cm} =\hbar^d  \int dy \, e^{-i\Pi\cdot y}\,
 _{\hat R}\langle R+\frac{\hbar}{2} y| \hat A |R-\frac{\hbar}{2} y\rangle_{\hat R}
\label{Eq2.7}
\end{eqnarray}
where the Fourier transform is taken in the usual generalized way\footnote{The Fourier transform $T_F$ of a generalized function $B \in {\cal S}'(\bkR^n)$ (for $n \ge 1$) is another generalized function $T_F[B]\in {\cal S}'(\bkR^n)$ which is defined by $ \langle T_F[B],t\rangle=\langle B,T_F[t]\rangle$ for all $t \in {\cal S}(\bkR^n)$ \cite{Hormander,Zemanian} and where $\langle A,t\rangle$ denotes the action of a distribution $A \in {\cal S}'(\bkR^n)$ on the test function $t \in {\cal S}(\bkR^n)$.} and the second form of the Weyl-Wigner map in terms of Dirac's bra and ket notation is more standard.
Note that we have introduced the subscript $\hat R$  to specify which eigenstates we are referring to. This will be relevant in the sequel.

There are two important restrictions of $W_{\xi}$:

(1) The first one is to the vector space $\hat{\cal F}$ of Hilbert-Schmidt operators on ${\cal H}$, which admit a representation of the form (\ref{Eq2.6}) with $A_K(R,R') \in {\cal F}$, regarded as
an algebra with respect to the standard operator product, which is an inner operation in $\hat{\cal F}$.
In this space we may also introduce the inner product $(\hat A,\hat B)_{\hat{\cal F}} \equiv tr(\hat A^{\dagger} \hat B)$ and the Weyl-Wigner map
$ W_{\xi}: \hat{\cal F} \longrightarrow {\cal F}$ becomes a one-to-one invertible unitary transformation.

(2) The second one is to the enveloping algebra $\hat{\cal A}({\cal H})$ of the Heisenberg-Weyl
Lie algebra which contains all polynomials of the fundamental operators $\hat R, \hat{\Pi}$ and $\hat{I}$
modulo the ideal generated by the Heisenberg commutation relations. In this case the Weyl-Wigner
transform $W_{\xi}:\hat{\cal A}({\cal H}) \longrightarrow {\cal A}(\bkR^{2d})$ becomes a one-to-one invertible map from $\hat{\cal A}({\cal H})$ to the algebra ${\cal A}(\bkR^{2d})$ of polynomial
functions on $\bkR^{2d}$. In particular $W_{\xi}(\hat I)=1$, $W_{\xi}(\hat R)=R$ and $W_{\xi}(\hat{\Pi})=\Pi$.\\

\noindent
The previous restrictions can be promoted to isomorphisms, if ${\cal F}$ and ${\cal A}(\bkR^{2d})$
are endowed with a suitable product. This is defined by:
\begin{equation}
W_{\xi}(\hat A) \star_{\xi} W_{\xi}(\hat B) \equiv W_{\xi}(\hat A \hat B)
\label{Eq2.8}
\end{equation}
for $\hat A,\hat B \in \hat{\cal S}'$ such that $\hat A\hat B \in \hat{\cal S}'$.
The $\star$-product admits the kernel representation:
\begin{eqnarray}
A(R,\Pi) \star_{\xi} B(R,\Pi) &=& {1\over(\pi \hbar)^{2d}} \int dR'\int dR'' \int d \Pi' \int d \Pi'' A(R', \Pi') B (R'', \Pi'') \times\nonumber\\
& \times & e^{{ - \frac{2i}{\hbar}\left[\Pi''\cdot(R-R')+\Pi\cdot(R'-R'')+\Pi'\cdot(R''-R)\right]}}
\label{Eq2.9}
\end{eqnarray}
and is an inner operation in ${\cal F}$ as well as in ${\cal A}(\bkR^{2d})$. The previous formula is also valid if we want to compute $A \star_{\xi} B$ with $A \in {\cal F}$ and $B\in {\cal A}(\bkR^{2d})$, in which case $A \star_{\xi} B \in {\cal S}' (\bkR^{2d})$. On the other hand, if $A \in {\cal A}(\bkR^{2d})$ and $B \in {\cal A}(\bkR^{2d}) \cup {\cal F}$ the $\star$-product can also be written in the well-known form \cite{Moyal,Groenewold}:
\begin{equation}
A(\xi) \star_{\xi} B(\xi) = A(\xi) e^{ \frac{i\hbar}{2} \buildrel{\leftarrow}\over\partial_{\xi_{\alpha}} J_{\alpha \beta} \buildrel{\rightarrow}\over\partial_{\xi_{\beta}}} B(\xi),
\label{Eq2.10}
\end{equation}
where $\partial_{\xi_{\alpha}} A(\xi) = \frac{\partial}{\partial_{\xi_{\alpha}}} A (\xi)$ and $J_{\alpha \beta}$ are the components of the sympletic matrix:
\begin{equation}
{\bf J} = \left(
\begin{array}{c c}
{\bf 0}_{d \times d} & {\bf I}_{d \times d}\\
- {\bf I}_{d \times d} & {\bf 0}_{d \times d}
\end{array}
\right).
\label{Eq2.11}
\end{equation}
Here ${\bf I}_{d \times d}$ denotes the ($d \times d$) identity matrix.\\

\noindent
The map Eq. (\ref{Eq2.7}) yields a phase space representation of quantum mechanics which, is by no means unique.
For instance the restriction $W_{\xi}:\hat{\cal A}({\cal H}) \longrightarrow {\cal A}(\bkR^{2d})$ is one of infinitely many isomorphisms relating the algebras $\hat{\cal A}({\cal H})$ and ${\cal A}(\bkR^{2d})$. This map is associated to the symmetric ordering prescription, the Weyl rule, for operators and admits a
large class of generalizations which are associated to different ordering prescriptions \cite{Baker}.
Moreover, even within the context of the Weyl rule, we may consider canonical operator transformations of the form:
\begin{equation}
\hat T: \hat{\cal S}' \longrightarrow \hat{\cal S}' ; \quad \hat A(\hat{\xi}) \longrightarrow \hat T[\hat A(\hat{\xi})] = \hat A'(\hat{\xi}')=\hat A(\hat{\xi}(\hat{\xi}'))~,
\label{Eq2.12}
\end{equation}
where $\hat{\xi}'=(\hat R',\hat{\Pi}')$ is a new set of fundamental Heisenberg variables and the action of $\hat T$
is tantamount to a change of representation. For the new set of fundamental variables we may write a new Weyl-Wigner map:
\begin{eqnarray}
W_{\xi'}: \hat{\cal S}' & \longrightarrow & {\cal S}'(\bkR^{2d}); \nonumber \\
\hat A & \longrightarrow & \tilde A (R',\Pi')= W_{\xi'}(\hat A) = \hbar^d  \int dy \, e^{-i\Pi'\cdot y} \,
_{\hat R'}\langle R'+\frac{\hbar}{2} y| \hat A |R'-\frac{\hbar}{2} y\rangle_{\hat R'}
\label{Eq2.13}
\end{eqnarray}
where now $|R' \pm \frac{\hbar}{2} y\rangle_{\hat R'}$ are eigenstates of $\hat{R}'$.
This map is not equivalent to the original one, $W_{\xi}$, given by Eq. (\ref{Eq2.7}). In fact, if we construct the coordinate transformation $\xi=\xi(\xi')=W_{\xi'}[\hat{\xi}(\hat{\xi}')]$ then, in general, $\tilde A (\xi')\not= A(\xi(\xi'))$ \cite{Dias3} and there is no trivial relation between $A$ and $\tilde A$. Notice that in Eq. (\ref{Eq2.13}), and also in Eq. (\ref{Eq2.7}), we did not specify whether $\hat A= \hat A(\hat{\xi})$ or $\hat A=\hat A'(\hat{\xi'})$. This is because $F(R',y)\equiv_{\hat R'}\langle R'+\frac{\hbar}{2} y| \hat A |R'-\frac{\hbar}{2} y\rangle_{\hat R'}$ is invariant under a change of representation, i.e.:
\begin{equation}
_{\hat R'}\langle R'+\frac{\hbar}{2} y| \hat A(\hat{\xi}) |R'-\frac{\hbar}{2} y\rangle_{\hat R'}=_{\hat R'}\langle R'+\frac{\hbar}{2} y| \hat A'(\hat{\xi}') |R'-\frac{\hbar}{2} y\rangle_{\hat R'}~.
\label{Eq2.14}
\end{equation}

We conclude that the following diagram does not close:
$$
\begin{array}{l c c}
\hat{A}(\hat{\xi}) & ------------- \longrightarrow & \hat{A}'(\hat{\xi}')= \hat A(\hat{\xi}(\hat{\xi}'))\\
& \hat T &\\
& & \\
W_{\xi} \downarrow & & \downarrow W_{\xi'}\\
& & \\
A(\xi)& & \tilde A (\xi')\\
& & \\
& \mbox{{\bf Diagram 1}} &
\end{array}
$$
and that, in general, the two maps $W_{\xi}$ and $W_{\xi'}$ yield two distinct phase space representations.\\

\noindent
Also arising from the Weyl rule, there is another set of quantum phase space representations
of the Heisenberg algebra. To construct them explicitly let us consider the more general case
where the transformation $\hat T$ (Eq. (\ref{Eq2.12})) is not required to be canonical and thus the new set of variables $\hat{\xi}'$ may no longer satisfy the Heisenberg commutation relations. In this case we shall
designate the new variables by $\hat z=(\hat q,\hat p)$. In general, there is no well-defined map $W_{z}$.
We demand that the transformation $\hat T$ be such that $\xi \longrightarrow z=z(\xi)=W_{\xi}(\hat z)$ is
a phase space diffeomorphism and define the classical counterpart of $\hat T$ by letting it to act as a coordinate transformation\footnote{Notice that we have used the notation $\tilde A$ in Eq. (\ref{Eq2.13}) and we use the notation $A'$ here. We shall reserve the notation with the prime for the case where the two objects are related by a coordinate transformation: $A' (z) = A (\xi (z))$.}:
\begin{equation}
T:{\cal S}'(\bkR^{2d})  \longrightarrow  {\cal S}'(\bkR^{2d}), \quad A(\xi) \longrightarrow T[A(\xi)]= A'(z)=A(\xi(z))~,
\label{Eq2.15}
\end{equation}
where $\xi(z)$ is the inverse function of $z(\xi)$. We may now consider the diagram:
$$
\begin{array}{l c c}
\hat{A}(\hat{\xi}) & ------------- \longrightarrow & \hat{A}'(\hat z)= \hat A(\hat{\xi}(\hat z))\\
& \hat T &\\
& & \\
W_{\xi} \downarrow & & \downarrow W^{\xi}_{z}\\
& & \\
A(\xi)&------------- \longrightarrow & A' (z)=A(\xi(z))\\
& T &\\
& & \\
& \mbox{{\bf Diagram 2}} &
\end{array}
$$
The generalized Weyl-Wigner map $W_{z}^{\xi}$ is the covariant generalization of $W_{\xi}$.
We have $W^{\xi}_{z}=T\circ W_{\xi} \circ \hat T^{-1}$ or more precisely \cite{Dias3}:
\begin{eqnarray}
W^{\xi}_{z}: \hat{\cal S}' & \longrightarrow & {\cal S}'(\bkR^{2d});  \nonumber\\
\hat A & \longrightarrow & A'(z)= W^{\xi}_{z}(\hat A) \nonumber\\
&&\hspace{1.1 cm} =\hbar^d \int dx \int dy \, e^{-i\Pi(z)\cdot y}
\delta (x-R(z)) \,_{\hat R}\langle x+\frac{\hbar}{2} y| \hat A |x-\frac{\hbar}{2} y\rangle_{\hat R}~,
\label{Eq2.16}
\end{eqnarray}
where again $|x \pm \frac{\hbar}{2} y\rangle_{\hat R}$ are eigenstates of $\hat{R}$. The map $W^{\xi}_{z}$
is still a one-to-one invertible map between the spaces $\hat{\cal S}'({\cal H})$ and ${\cal S}'(\bkR^{2d})$ and even if $z$ is a second set of canonical coordinates, in general, we have $W^{\xi}_{z} \not=W_z$. The map $W^{\xi}_{z}$ can be applied both to a general observable $\hat A \in \hat{\cal A}({\cal H})$ as well as to the
density matrix $\hat{\rho}(t)=|\psi(t)\rangle\langle\psi(t)| \in \hat{\cal F}$. In the first case it yields
the $^{\xi}_{z}$-Weyl symbol $A'(z)=W^{\xi}_{z}(\hat A)$ of the original quantum operator and,
in the second case, the covariant generalization of the celebrated Wigner
function $f^{W'}(z,t)=\frac{1}{(2\pi \hbar)^d}W^{\xi}_{z}(\hat{\rho}(t))$ \cite{Wigner}.
Moreover, the
two restrictions $W^{\xi}_{z}:\hat{\cal F} \longrightarrow {\cal F}$
and $W^{\xi}_{z}:\hat{\cal A}({\cal H}) \longrightarrow {\cal A}(\bkR^{2d})$
can be turned into isomorphisms of Lie algebras once a suitable $\star$-product
is introduced in ${\cal F}$ and ${\cal A}(\bkR^{2d})$.\\

\noindent
In analogy with Eq. (\ref{Eq2.8}) the covariant star-product ${\star}_{z}$ and Moyal bracket $[,]_{z}$ are defined by \cite{Bayen,Vey,Dias3}:
\begin{eqnarray}
W^{\xi}_{z}(\hat A){\star}_{z} W^{\xi}_{z}(\hat B) & \equiv &  W^{\xi}_{z}(\hat A\hat B) \nonumber\\
\left[ W^{\xi}_{z}(\hat A),W^{\xi}_{z}(\hat B) \right]_{z}  & \equiv &  W^{\xi}_{z}({1\over i\hbar}[\hat A,\hat B]) = {1\over i\hbar} (W^{\xi}_{z}(\hat A){\star}_{z} W^{\xi}_{z}(\hat B)-W^{\xi}_{z}(\hat B){\star}_{z} W^{\xi}_{z}(\hat A))
\label{Eq2.17}
\end{eqnarray}
for all $\hat A,\hat B\in \hat{\cal S}'$ such that $\hat A\hat B,\hat B\hat A \in \hat{\cal S}'$. From its definition it follows  that the $\star_z$-product satisfies:
\begin{equation}
A'(z) {\star}_{z} B'(z) = A(\xi(z)){\star}_{\xi} B(\xi(z))
\label{Eq2.18}
\end{equation}
and that for $A, B \in {\cal A}(\bkR^{2d})$ it can be written in the explicit form\footnote{Notice that
the covariant version of the ${\star}$-product for the configuration variables has been considered in the context of field theory
in order to minimally couple a scalar field with gravity \cite{BGuisado1}.}:
\begin{equation}
A'(z) \star_z B'(z) = A'(z)  e^{\frac{i\hbar}{2} \buildrel{\leftarrow}\over\nabla_{z_{\alpha}} \Omega_{\alpha \beta} \buildrel{\rightarrow}\over\nabla_{z_{\beta}}} B'(z).
\label{Eq2.19}
\end{equation}
This formula truncates at some finite order and is also valid for the case where $A$ (or $B$) $\in {\cal A}(\bkR^{2d})$ and $B$ (or $A$) $ \in {\cal F}$. In Eq. (\ref{Eq2.19}) the covariant derivatives are given by:
\begin{eqnarray}
\nabla_{z_{\alpha}} A'(z) &=& \partial_{z_{\alpha}} A'(z) = \frac{\partial}{\partial_{z_{\alpha}}} A' (z) \hspace{1 cm}\nonumber\\
\nabla_{z_{\alpha}} \nabla_{z_{\beta}} A'(z) &=& \partial_{z_{\alpha}} \partial_{z_{\beta}} A'(z) - \Gamma_{\alpha \beta \gamma} \partial_{z_{\gamma}} A'(z) \hspace{1 cm},
\label{Eq2.20}
\end{eqnarray}
where the Christoffel symbols read (we assume that the phase space has a flat structure):
\begin{equation}
\Gamma_{\alpha \beta \gamma} = {\partial z_{\gamma}\over\partial \xi_{\sigma}}{\partial^2 \xi_{\sigma}\over\partial z_{\alpha} \partial z_{\beta}}.
\label{Eq2.21}
\end{equation}
and the transformed symplectic form ${\bf \Omega}$ is given by:
\begin{equation}
\Omega_{\alpha \beta} =  {\partial z_{\alpha}\over\partial \xi_{\gamma}}{\partial z_{\beta}\over\partial \xi_{\sigma}} J_{\gamma \sigma} =  {\partial z_{\alpha}\over\partial R_k} {\partial z_{\beta}\over\partial \Pi_k} - {\partial z_{\alpha}\over\partial \Pi_k} {\partial z_{\beta}\over\partial R_k}.
\label{Eq2.22}
\end{equation}
We will see in the next section that when the transformation $\xi \longrightarrow \xi(z)$ is linear, the kernel expression of the $\star$-product Eq. (\ref{Eq2.9}) also admits a simple covariant generalization.

We complete the presentation of the main structures of the covariant formulation of phase space quantum mechanics by introducing the trace formula for a general trace-class operator $\hat A \in \hat{\cal F}$ \cite{Dias3}:
\begin{equation}
Tr(\hat A) = {1\over(2\pi \hbar)^{d}} \int d z \left|{\partial \xi\over\partial z} \right| W^{\xi}_{z}(\hat A)
= {1\over(2\pi \hbar)^{d}} \int  {d z\over\sqrt{\det {\bf \Omega}}}  A'(z),
\label{Eq2.23}
\end{equation}
and for the product of two operators $\hat A,\hat B \in \hat{\cal F}$ we get:
\begin{equation}
Tr(\hat A \hat B) = {1\over(2\pi \hbar)^{d}} \int  {d z\over\sqrt{\det {\bf \Omega}}}  A'(z) \star_z B'(z)=
{1\over(2\pi \hbar)^{d}} \int {d z\over\sqrt{\det {\bf \Omega}}}  A'(z)B'(z),
\label{Eq2.24}
\end{equation}
where we have used the fact that (cf. Eq. (\ref{Eq2.18})):
\begin{eqnarray}
\int  {d z\over\sqrt{\det {\bf \Omega}}}  A'(z) \star_z B'(z)&=& \int d\xi \,  A(\xi) \star_{\xi} B(\xi) \nonumber\\
&=&\int d\xi \, A(\xi) B(\xi) \nonumber\\
&=&\int  {d z\over\sqrt{\det {\bf \Omega}}}  A'(z)B'(z).
\label{Eq2.25}
\end{eqnarray}
It follows that:
\begin{equation}
(\hat A,\hat B)_{\hat{\cal F}}=tr(\hat A^{\dagger} \hat B)= {1\over(2\pi \hbar)^{d}} \int  {d z\over\sqrt{\det {\bf \Omega}}} \overline{ A'(z)} B'(z) =(A',B')_{\cal F}
\label{Eq2.26}
\end{equation}
and so $W^{\xi}_{z}:\hat{\cal F} \longrightarrow {\cal F}$ is an unitary transformation.\\

\noindent
When formulated in terms of these structures phase space quantum mechanics becomes fully
invariant under the action of general coordinate transformations.
The covariant generalization of the Moyal and stargenvalue equations read:
\begin{equation}
{\partial f^{W'} \over \partial t} = [H',f^{W'}]_{z}\hspace{0.5 cm},\hspace{0.5 cm} 
\left\{ \begin{array}{lll}
A'(z) {\star}_{z} \rho'_{a,b}(z) & = &  a \rho'_{a,b}(z) \\
\rho'_{a,b}(z) {\star}_{z} A'(z) & = &  b \rho'_{a,b}(z)
\end{array} \right. 
\label{Eq2.27}
\end{equation}
where $\rho'_{a,b}(z)$ is the $a$-left and $b$-right ${\star}_{z}$-genfunction
of $A'(z)$, $\rho'_{a,b}(z) = (2 \pi \hbar )^{-d} W_z^{\xi} (|a\rangle\langle b|)$ and $H'$
is the $_z^{\xi}$-symbol of the Hamiltonian $\hat H$. Finally, the probability distributions for a general observable is given by:
\begin{equation}
P(A'(z,t)=a)=  \int {d z\over\sqrt{\det {\bf \Omega}}}
\rho'_{a,a}(z)  f^{W'}(z,t)\hspace{0.1cm}.
\label{Eq2.28}
\end{equation}

A final remark is in order: from the trace formula, it follows that the covariant
Wigner function $f^{W'}(z,t)=\frac{1}{(2\pi \hbar)^d}W^{\xi}_{z}(\hat{\rho}(t))$ does not
satisfy the normalization condition $\int d z f^{W'}(z,t) =1$. Instead we have:
\begin{equation}
\int {d z\over \sqrt{\det {\bf \Omega}}} f^{W'}(z,t) =1
\label{Eq2.29}
\end{equation}
In order to prevent this situation, we shall introduce the following convention, which
is the one that renders our future results simpler. If $\det {\bf \Omega}$ is a
constant, which is the case for $T$ the SW map, we redefine the Wigner function as:
\begin{equation}
f^{W'}(z,t) \longrightarrow {1\over\sqrt{\det {\bf \Omega}}} f^{W'}(z,t)
\label{Eq2.30}
\end{equation}
so that the quasi-distribution is normalized. On the other hand if $\det {\bf \Omega}$ is not
a constant, then we shall keep the original definition in order not to be forced to change the main
equations that should be satisfied by the Wigner function, such as the Moyal and stargenvalue equations.

\section{Noncommutative quantum mechanics in phase space}

In this section we shall consider a quantum system in $d$ dimensions with both spatial and momentum noncommutativity. This is expressed in terms of the extended Heisenberg algebra satisfying the commutation relations:
\begin{equation}
\left[\hat q_i, \hat q_j \right] = i \theta_{ij} , \hspace{0.5 cm} \left[\hat q_i, \hat p_j \right] = i \hbar \delta_{ij} ,
\hspace{0.5 cm} \left[\hat p_i, \hat p_j \right] = i \eta_{ij} ,  \hspace{0.5 cm} i,j= 1, ... ,d
\label{Eq3.1}
\end{equation}
where $\eta_{ij}$ and $\theta_{ij}$ are antisymmetric real constant ($d \times d$) matrices.
We shall assume that these matrices are both invertible and that
\begin{equation}
\Sigma_{ij} \equiv \delta_{ij} + {1\over\hbar^2}  \theta_{ik} \eta_{kj}
\label{Eq3.2}
\end{equation}
is equally an invertible matrix. This will certainly happen, if for any matrix elements $\eta$ and
$\theta$, their product is considerably smaller than $\hbar^2$:
\begin{equation}
\theta \eta << \hbar^2.
\label{Eq3.3}
\end{equation}
We will tacitly assume this to be the case. Notice that this condition is experimentally fulfilled, for instance, in the case of the noncommutative gravitational quantum well \cite{Bertolami1}.

\noindent It is well-known that under a linear transformation this algebra can be mapped
to the usual Heisenberg algebra:
\begin{equation}
\left[\hat R_i, \hat R_j \right] = 0 , \hspace{0.5 cm} \left[\hat R_i, \hat \Pi_j \right]
= i \hbar \delta_{ij} , \hspace{0.5 cm} \left[\hat \Pi_i, \hat \Pi_j \right] = 0 ,
\hspace{0.5 cm} i,j= 1, ... ,d ~, \label{Eq3.4}
\end{equation}
via the SW map \cite{Seiberg}, which can be cast in the form:
\begin{equation}
\hat q_i = A_{ij} \hat R_j + B_{ij} \hat \Pi_j \hspace{1 cm}
\hat p_i = C_{ij} \hat R_j + D_{ij} \hat \Pi_j
\label{Eq3.5}
\end{equation}
where ${\bf A},{\bf B},{\bf C},{\bf D}$ are real constant matrices. We shall assume that this transformation
is invertible. The transformation is easily shown to obey the following lemma \cite{Rosenbaum2}.

\vspace{0.3 cm}
\noindent
{\bf Lemma 3.1:} The matrices ${\bf A},{\bf B},{\bf C},{\bf D}$ are solutions of the equations
\begin{equation}
{\bf A} {\bf D}^T - {\bf B} {\bf C}^T = {\bf I}_{d \times d} \hspace{1 cm} {\bf A} {\bf B}^T - {\bf B} {\bf A}^T = {1\over\hbar} {\bf \Theta} \hspace{1 cm}
{\bf C} {\bf D}^T - {\bf D} {\bf C}^T = {1\over\hbar} {\bf N}~,
\label{Eq3.6}
\end{equation}
where the superscript ``T" stands for matrix transposition. ${\bf A}, {\bf B}, {\bf C}, {\bf D}, {\bf \Theta}, {\bf N}$ are the matrices with entries $A_{ij}, B_{ij}, C_{ij}, D_{ij} , \theta_{ij} , \eta_{ij}$, respectively. 

\vspace{0.3 cm}
\noindent
Thanks to this linear transformation, the noncommutative algebra Eq. (\ref{Eq3.1})
admits a representation in terms of the Hilbert space of ordinary quantum mechanics.
Notice that the SW map is not unique. Indeed, the Heisenberg algebra (\ref{Eq3.4}) is
only defined up to a unitary transformation. Consequently, there should be an infinity
set of solutions to the set of constraints (\ref{Eq3.6}). Indeed, there are $4$ $(d \times d)$
matrices to be determined, i.e. $4 d^2$ real parameters. However, in (\ref{Eq3.6})
there are $d(3d-1)/2$ independent equations, leaving a total of $d(5d+1)/2$ free parameters.

The aim of this section is to find a phase space representation of the commutation relations (\ref{Eq3.1}) and, by doing so, provide a phase space formulation of noncommutative quantum mechanics.
Let $\hat A(\hat q,\hat p), \hat B(\hat q,\hat p) \in \hat{\cal S}'$ be two generalized
operators. Two important
subspaces of $\hat{\cal S}'$ are the enveloping algebra of the extended Heisenberg
algebra $\hat{\cal A}_E({\cal H})$ and the algebra of Hilbert-Schmidt operators $\hat{\cal F}$. It follows from the SW map that $\hat{\cal A}_E({\cal H}) =\hat{\cal A}({\cal H})$.
We are looking for a one-to-one invertible linear map $V:\hat{\cal S}' \longrightarrow {\cal S}'(\bkR^{2d})$ satisfying:

(i) $V(\hat I)= 1$

(ii) $V(\hat q)=q$

(iii) $V(\hat p)=p$

(iv) $V\left(\hat A(\hat q,\hat p) \hat B(\hat q,\hat p) \right)= V \left(\hat A(\hat q,\hat p) \right) \star V\left(\hat B(\hat q,\hat p) \right)$, for some suitable product $\star$ and provided $\hat A \hat B \in \hat{\cal S}'$.

Notice that these properties immediately impose the extended Heisenberg commutation relations in phase space:
\begin{equation}
V \left([\hat z_{\alpha}, \hat z_{\beta}] \right)= V(\hat z_{\alpha}) \star V(\hat z_{\beta})-V(\hat z_{\beta}) \star V(\hat z_{\alpha}) =[V(\hat z_{\alpha}),V(\hat z_{\alpha})]_{\star}
\label{Eq3.1.1}
\end{equation}
where we defined the bracket associated to the product $\star$.

There is a large set of maps $V$ that satisfy properties (i) to (iv). This set can be divided into
several subsets, each one being characterized by a particular ordering prescription for operators.
The class of maps $V$ which are associated with the Weyl rule can be constructed by resorting to the standard Weyl-Wigner map:
\begin{equation}
\underbrace{\begin{array}{ccccccc}
\hat A' (\hat z) &\longrightarrow & \hat A(\hat{\xi}) & \longrightarrow & A(\xi) & \longrightarrow & A'(z) \\
& \hat{SW} & & W_{\xi} & & SW^{-1} &
\end{array}}_{W_z^{\xi}}
\label{Eq3.1.2}
\end{equation}
and since the SW transformation is linear, there are no ordering ambiguities and thus $\hat{SW}$ and $SW$ have the same functional form.
One can easily see that $W_z^{\xi}$ provides the desired phase space representation for NCQM.
Indeed the map $W_z^{\xi}$ satisfies the four required properties and defines a new $\star$-product
in phase space through condition (iv). In section 3-A and 3-B we will study the algebraic structure of the new theory and in section 3-C we shall characterize its states. Finally, in section 3-D we shall also show that the
map $W_z^{\xi}$, and thus the entire structure of the phase space representation of NCQM, is independent of the particular choice for the SW map.

\subsection{The $\star$ product}

Using the generalized Weyl-Wigner map (\ref{Eq3.1.2}) we now construct a $\star$-product for the NCQM in phase space. In ${\cal A}(\bkR^{2d})$ we can use Eq. (\ref{Eq2.19}) to write the $\star$-product in terms of the variables $z$. In this context
in order to keep the notation simple, we shall write $A(z)$ and $\star$
instead of $A'(z)$ and $\star_z$, respectively. Since the SW map is a linear transformation, the
Christoffel symbols in (\ref{Eq2.21}) vanish and the covariant derivatives (\ref{Eq2.20}) reduce to
the standard derivatives. As for the symplectic form, we get from Eqs. (\ref{Eq2.22}), (\ref{Eq3.5}) and (\ref{Eq3.6}):
\begin{equation}
\Omega_{q_i q_j} = A_{ik}B_{jk} - B_{ik}A_{jk} = {1\over\hbar} \theta_{ij}.
\label{Eq3.1.3}
\end{equation}
Likewise:
\begin{equation}
\Omega_{q_i p_j} = \delta_{ij}\hspace{0.5 cm} , \hspace{0.5 cm} \Omega_{p_i p_j} = {1\over\hbar} \eta_{ij}.
\label{Eq3.1.4}
\end{equation}
Altogether:
\begin{equation}
{\bf \Omega} = \left(
\begin{array}{c c}
\frac{1}{\hbar} {\bf \Theta} & {\bf I}_{d \times d}\\
- {\bf I}_{d \times d} &  \frac{1}{\hbar} {\bf N}
\end{array}
\right).
\label{Eq3.1.5}
\end{equation}
We have thus proved the following theorem:

\vspace{0.3 cm}
\noindent
{\bf Theorem 3.2:}
For $A,B \in {\cal A} (\bkR^{2d})$ the $\star$-product in the noncommutative variables $z=(q,p)$ reads:
\begin{equation}
A(z) \star B(z) = A(z) e^{\frac{i \hbar}{2} \buildrel{\leftarrow}\over\partial_{z_{\alpha}}  \Omega_{\alpha \beta} \buildrel{\rightarrow}\over\partial_{z_{\beta}} }  B(z) = A(z) {\star}_{\hbar} {\star}_{\theta} {\star}_{\eta} B(z),
\label{Eq3.1.6}
\end{equation}
where
\begin{eqnarray}
A(z) {\star}_{\hbar} B(z) \equiv A (z) e^{\frac{i \hbar}{2} \buildrel{\leftarrow}\over\partial_{z_{\alpha}} J_{\alpha \beta}  \buildrel{\rightarrow}\over\partial_{z_{\beta}}} B (z) \label{Eq3.1.7.A}\\
A(z) {\star}_{\theta} B(z) \equiv  A(z) e^{\frac{i}{2} \frac{\buildrel{\leftarrow}\over\partial}{\partial q_i} \theta_{ij} \frac{\buildrel{\rightarrow}\over\partial}{\partial q_j}} B(z) \label{Eq3.1.7.B}\\
A(z) {\star}_{\eta} B(z) \equiv A(z) e^{\frac{i}{2} \frac{\buildrel{\leftarrow}\over\partial}{\partial p_i} \eta_{ij} \frac{\buildrel{\rightarrow}\over\partial}{\partial p_j}} B(z),
\label{Eq3.1.7.C}
\end{eqnarray}
We can cast all these products in a compact notation:
\begin{equation}
A(\chi) \star_{\Lambda} B(\chi) = A(\chi) e^{\frac{i}{2} \buildrel{\leftarrow}\over\partial_{\chi_R}  \Lambda_{RS} \buildrel{\rightarrow}\over\partial_{\chi_S} }  B(\chi)~.
\label{Eq3.1.8}
\end{equation}
In the case of $\star$ and $\star_{\hbar}$ the variable $\chi$ stands for $\chi=z=(q,p)$ in the $2d$-dimensional phase space $(R,S=1, \cdots, 2d)$. The symplectic matrices in these cases read:
\begin{equation}
\begin{array}{l l l l l}
{\bf \Lambda} = \hbar {\bf \Omega}, & \mbox{if } \star_{\Lambda} = \star & \hspace{0.5 cm} &,\hspace{0.5cm}
{\bf \Lambda} = \hbar {\bf J}, & \mbox{if } \star_{\Lambda} = \star_{\hbar}~.
\end{array}
\label{Eq3.1.9}
\end{equation}
In the case of $\star_{\theta}$ and $\star_{\eta}$ the variable $\chi$ stands for $\chi=q$ or $\chi=p$ $ (R,S=1, \cdots, d)$, respectively,
and the remaining variables in Eqs. (\ref{Eq3.1.7.B}) and (\ref{Eq3.1.7.C}), $p$ or $q$, are just regarded as fixed parameters in the corresponding $d$-dimensional configuration or momentum spaces. The symplectic matrices thus read:
\begin{equation}
\begin{array}{l l l l l}
{\bf \Lambda} = {\bf \Theta}, & \mbox{if } \star_{\Lambda} = \star_{\theta} & \hspace{0.5 cm} &,\hspace{0.5cm}
{\bf \Lambda} = {\bf N}, & \mbox{if } \star_{\Lambda} = \star_{\eta}~.
\end{array}
\label{Eq3.1.10}
\end{equation}

\noindent
Then we can derive two additional representations for these $\star$-products.

\vspace{0.3 cm}
\noindent
{\bf Lemma 3.3:} A $\star$-product of the form (\ref{Eq3.1.8}) acting on the space of polynomials on phase space ($\star_{\Lambda} = \star$ or $\star_{\Lambda} = \star_{\hbar}$),
configuration space ($\star_{\Lambda} = \star_{\theta}$) or momentum space ($\star_{\Lambda} = \star_{\eta}$), can be represented as a Bopp shift:
\begin{equation}
A(\chi) \star_{\Lambda} B(\chi) = A \left(\chi + \frac{i}{2} {\bf \Lambda} \buildrel{\rightarrow}\over\partial_{\chi} \right) B(\chi) = A(\chi) B \left(\chi - \frac{i}{2} {\bf \Lambda} \buildrel{\leftarrow}\over\partial_{\chi} \right).
\label{Eq3.1.11}
\end{equation}

\vspace{0.3 cm}
\noindent
{\bf Proof:} Assuming that $A(\chi)$, $B(\chi)$ are polynomials, then they admit a generalized Fourier transform. We may thus write:
\begin{equation}
A(\chi) = \int d \zeta \hspace{0.2 cm} f (\zeta ) e^{i \zeta \cdot \chi}\hspace{0.5 cm}, \hspace{0.5 cm}
B(\chi) = \int d \sigma \hspace{0.2 cm} g (\sigma ) e^{i \sigma \cdot \chi}~.
\label{Eq3.1.12}
\end{equation}
Here $\chi= z$ (for $\star_{\Lambda} = \star$ or $\star_{\Lambda}= \star_{\hbar}$), $\chi=q$ (for $\star_{\Lambda} = \star_{\theta}$) or $\chi=p$ (for $\star_{\Lambda} = \star_{\eta}$). The coefficients $f(\zeta), g(\sigma)$ are singular. From (\ref{Eq3.1.8}) we get:
\begin{eqnarray}
A(\chi) \star_{\Lambda} B(\chi) &=& \int d \zeta \int d \sigma \hspace{0.2 cm} f (\zeta ) g (\sigma ) e^{i \zeta \cdot \chi} e^{\frac{i}{2} \buildrel{\leftarrow}\over\partial_{\chi_R}  \Lambda_{RS} \buildrel{\rightarrow}\over\partial_{\chi_S} } e^{i \sigma \cdot \chi}\nonumber\\
&=& \int d \zeta \int d \sigma \hspace{0.2 cm} f (\zeta ) g (\sigma ) e^{i \zeta_R \left( \chi_R + \frac{i}{2} \Lambda_{RS}\buildrel{\rightarrow}\over\partial_{\chi_S} \right) } e^{i \sigma \cdot \chi}\nonumber\\
&=& A \left(\chi + \frac{i}{2} {\bf \Lambda} \buildrel{\rightarrow}\over\partial_{\chi} \right) B(\chi)~.
\label{Eq3.1.13}
\end{eqnarray}
The second identity in (\ref{Eq3.1.11}) is proved in a similar way.$_{\Box}$

\vspace{0.3 cm}
\noindent
{\bf Lemma 3.4:} Let the matrix ${\bf \Lambda}$ be invertible. Then the $\star$-product of the form (\ref{Eq3.1.8}) admits a kernel representation:
\begin{equation}
A(\chi) \star_{\Lambda} B(\chi) = {1\over\pi^n | \det {\bf \Lambda} |} \int d \chi' \int d \chi'' \hspace{0.2 cm} A(\chi') B(\chi'')e^{[2i (\chi- \chi')^T {\bf \Lambda}^{-1} (\chi''- \chi)]},
\label{Eq3.1.14}
\end{equation}
for $A, B \in {\cal A}(\bkR^n)$. Moreover, this kernel representation is equally valid for $A, B \in {\cal A}(\bkR^{2d}) \cup {\cal F}$ if $\Lambda = \hbar \Omega$. Here $n$ stands for $2d$ or $d$ depending on whether the $\star$-product is $\star$, $\star_{\hbar}$, or $\star_{\theta}$, $\star_{\eta}$, respectively.

\vspace{0.3 cm}
\noindent
{\bf Proof:} Let us prove the equivalence of (\ref{Eq3.1.8}) and (\ref{Eq3.1.14}) for polynomials $A,B$. Again, $A(\chi)$ and $B(\chi)$ admit the Fourier transforms (\ref{Eq3.1.12}) with inverse forms given by:
\begin{equation}
f( \zeta) = {1\over(2 \pi)^n} \int d \chi \hspace{0.2 cm} A( \chi) e^{-i \zeta \cdot \chi} \hspace{0.5 cm}, \hspace{0.5 cm} g( \sigma) = \frac{1}{(2 \pi)^n} \int d \chi \hspace{0.2 cm} B( \chi) e^{-i \sigma \cdot \chi}.
\label{Eq3.1.15}
\end{equation}
From (\ref{Eq3.1.11}) we get:
\begin{eqnarray}
A( \chi) \star_{\Lambda} B( \chi) &=& \int d \zeta \int d \sigma \hspace{0.2 cm} f (\zeta ) g (\sigma ) e^{i \zeta_R \left( \chi_R + \frac{i}{2} \Lambda_{RS}\buildrel{\rightarrow}\over\partial_{\chi_S} \right) } e^{i \sigma \cdot \chi} =\nonumber\\
& = & \int d \zeta \int d \sigma \hspace{0.2 cm} f (\zeta ) g (\sigma ) e^{\left(i \zeta \cdot \chi + i \sigma \cdot \chi - \frac{i}{2} \zeta^T {\bf \Lambda} \sigma \right)}.
\label{Eq3.1.16}
\end{eqnarray}
Upon substitution of Eq. (\ref{Eq3.1.15}) into the previous equation, we get:
\begin{eqnarray}
A( \chi) \star_{\Lambda} B( \chi) &=& {1\over(2 \pi)^{2n}} \int d \zeta \int d \sigma \int d \chi' \int d \chi'' A( \chi') B( \chi'') \times\nonumber\\
& \times & \exp \left(- i \zeta \cdot \chi' - i \sigma \cdot \chi''+ i \zeta \cdot \chi + i \sigma \cdot \chi - \frac{i}{2} \zeta^T {\bf \Lambda} \sigma \right).
\label{Eq3.1.17}
\end{eqnarray}
Let us first perform the integration over $\sigma$:
\begin{eqnarray}
\int d \sigma e^{(- i \sigma \cdot \chi''+  i \sigma \cdot \chi - \frac{i}{2} \zeta^T {\bf \Lambda} \sigma)} &=& (2 \pi)^n \delta \left(\chi - \chi'' - \frac{1}{2} \zeta^T {\bf \Lambda} \right) \nonumber\\
&=& {(4 \pi)^n\over | \det {\bf \Lambda} |} \delta \left( \zeta - 2 {\bf \Lambda}^{-1} ( \chi''- \chi) \right).
\label{Eq3.1.18}
\end{eqnarray}
From Eqs. (\ref{Eq3.1.17}), (\ref{Eq3.1.18}) we recover (\ref{Eq3.1.14}) upon integration over $\zeta$.

Finally, we prove the last statement in the lemma. Let $C(\xi )$ and $D (\xi)$ be the Weyl symbols of some operators $\hat C$, $\hat D$ in $\hat{{\cal A}} ({\cal H}) \cup \hat{{\cal F}}$. The kernel representation of the $\star_{\hbar}$-product Eq. (\ref{Eq2.9}) can be writen in the form:
\begin{equation}
C(\xi) \star_{\hbar} D( \xi) = {1\over(\pi \hbar)^{2d}} \int d \xi' \int d \xi'' \hspace{0.2 cm} C ( \xi') D (\xi'') e^{[ - \frac{2i}{\hbar} (\xi - \xi')^T {\bf J} (\xi'' - \xi)]}.
\label{Eq3.1.19}
\end{equation}
If we perform the SW transformation:
\begin{equation}
\xi \longrightarrow z = {\bf T} \xi \hspace{0.2 cm},\hspace{0.2cm} C(\xi) \longrightarrow C' (z) = C (\xi (z)) \hspace{0.2 cm},\hspace{0.2cm} D(\xi) \longrightarrow D' (z) = D( \xi (z)),
\label{Eq3.1.20}
\end{equation}
with $T_{\alpha \beta} = \frac{\partial z_{\alpha}}{\partial \xi_{\beta}}$. The symplectic matrix transforms as in (\ref{Eq2.22}):
\begin{equation}
{\bf \Omega} = {\bf T} {\bf J} {\bf T}^T.
\label{Eq3.1.21}
\end{equation}
and $\det {\bf T} = (\det {\bf \Omega} )^{1/2}$. Under the SW map the kernel representation (\ref{Eq3.1.19}) transforms according to (cf. (\ref{Eq2.18})):
\begin{eqnarray}
&&\left. C(\xi) \star_{\hbar} D (\xi)\right|_{\xi=\xi (z)} = C' (z) \star D' (z) \nonumber\\
&=&{1\over(\pi \hbar)^{2d}} \int d z' \int d z'' (\det {\bf T} )^{-2} C' (z') D' (z'') e^{[ - \frac{2i}{\hbar} (z - z')^T ({\bf T}^{-1})^T {\bf J} {\bf T}^{-1} (z'' - z)]}.
\label{Eq3.1.22}
\end{eqnarray}
From (\ref{Eq3.1.21}) we have ${\bf \Omega}^{-1} = - ( {\bf T}^{-1} )^T {\bf J} {\bf T}^{-1}$. And thus:
\begin{equation}
C' (z) \star D' (z) = {1\over(\pi \hbar)^{2d} |\det {\bf \Omega} |} \int d z' \int d z'' \hspace{0.2 cm} C' (z') D' (z'') e^{[ \frac{2i}{\hbar} (z - z')^T {\bf \Omega}^{-1} (z'' - z)]}.
\label{Eq3.1.23}
\end{equation}
If we redefine $A (z) \equiv C' (z)$, $B (z) \equiv D' (z)$ we recover (\ref{Eq3.1.14}).$_{\Box}$

\vspace{0.3 cm}
\noindent
{\bf Remmark 3.5:} Notice that for the expression (\ref{Eq3.1.14}) to make sense, it is crucial that the matrix ${\bf \Lambda}$ is invertible. According to our assumptions, Eq. (\ref{Eq3.3} and the preceding discussion, this is certainly the case for the matrices ${\bf J}$, ${\bf \Theta}$ and ${\bf \eta}$. The matrix ${\bf \Omega}$ also has an inverse:
\begin{equation}
{\bf \Omega}^{-1} = \left(
\begin{array}{c c}
\frac{1}{\hbar} {\bf N} {\bf \Sigma}^{-1}  & - \left({\bf \Sigma}^{-1} \right)^T \\
{\bf \Sigma}^{-1} & \frac{1}{\hbar} {\bf \Theta} \left({\bf \Sigma}^{-1} \right)^T
\end{array}
\right)
\label{Eq3.1.24}
\end{equation}
where ${\bf \Sigma}^{-1}$ is the inverse of the matrix (\ref{Eq3.2}).

\vspace{0.3 cm}
\noindent
{\bf Theorem 3.6:} For a $\star$-product of the form (\ref{Eq3.1.14}), the following identity holds:
\begin{equation}
\int d \chi \hspace{0.2 cm} A( \chi) \star_{\Lambda} B( \chi) = \int d \chi \hspace{0.2 cm} A( \chi)  B( \chi).
\label{Eq3.1.25}
\end{equation}

\vspace{0.3 cm}
\noindent
{\bf Proof:} From the kernel representation, we have:
\begin{eqnarray}
\int {d \chi A( \chi) \star_{\Lambda} B( \chi)} &=& {1\over\pi^n | \det {\bf \Lambda} |} \int d \chi  \int d \chi' \int d  \chi'' A( \chi') B( \chi'') e^{[2i ( \chi- \chi')^T {\bf \Lambda}^{-1} ( \chi''- \chi)]}  \nonumber\\
&=& {1\over\pi^n | \det {\bf \Lambda} |}  \int d \chi' \int d \chi'' \pi^n | \det  {\bf \Lambda}| \delta( \chi'-  \chi'') A( \chi') B( \chi'') e^{ - 2i \chi'^T {\bf \Lambda}^{-1}  \chi''}\nonumber\\
&=& \int {d \chi A( \chi)  B( \chi)}~,
\label{Eq3.1.26}
\end{eqnarray}
where the antisymmetry of ${\bf \Lambda}$ has been used in the last step. Notice that this formula is in agreement with (\ref{Eq2.24}).$_{\Box}$

\subsection{The Moyal bracket}

The extended Moyal bracket is defined according to Eq. (\ref{Eq2.17}).

\vspace{0.3 cm}
\noindent
{\bf Definition 3.7:} For $A,B \in {\cal S}'(\bkR^{2d})$ such that $A\star B$ and $B\star A$ are also in ${\cal S}'(\bkR^{2d})$ the noncommutative Moyal bracket is defined by:
\begin{equation}
\left[A (z) , B (z) \right]_{\star} \equiv {1\over i \hbar} \left( A (z) \star B (z) - B (z) \star A (z) \right)
\label{Eq3.2.1}
\end{equation}
In particular, this definition is valid for $A,B \in {\cal A} ( \bkR^{2d} ) \cup {\cal F}$.$_{\Box}$

\vspace{0.3 cm}
\noindent
For $A \in {\cal A} ( \bkR^{2d} )$ and $B \in {\cal A} (\bkR^{2d}) \cup {\cal F}$ one easily realizes that the Moyal bracket can be written in the form:
\begin{eqnarray}
\left[A (z) , B (z) \right]_{\star} &\equiv& {1\over i \hbar} \left( A (z) \star B (z) - B (z) \star A (z) \right)\nonumber\\
&=& {1\over i \hbar} \left( A(z) e^{\frac{i \hbar}{2} \buildrel{\leftarrow}\over\partial_{z_{\alpha}}  \Omega_{\alpha \beta} \buildrel{\rightarrow}\over\partial_{z_{\beta}}} B(z) - B(z) e^{\frac{i \hbar}{2} \buildrel{\leftarrow}\over\partial_{z_{\alpha}}  \Omega_{\alpha \beta} \buildrel{\rightarrow}\over\partial_{z_{\beta}}} A(z) \right) \nonumber\\
&=& {1\over i \hbar} A(z) \left(  e^{\frac{i \hbar}{2} \buildrel{\leftarrow}\over\partial_{z_{\alpha}}  \Omega_{\alpha \beta} \buildrel{\rightarrow}\over\partial_{z_{\beta}}}  -  e^{- \frac{i \hbar}{2} \buildrel{\leftarrow}\over\partial_{z_{\alpha}}  \Omega_{\alpha \beta} \buildrel{\rightarrow}\over\partial_{z_{\beta}}} \right) B(z) \nonumber\\
&=& {2\over \hbar} A(z) \sin \left( \frac{\hbar}{2} \buildrel{\leftarrow}\over\partial_{z_{\alpha}}  \Omega_{\alpha \beta} \buildrel{\rightarrow}\over\partial_{z_{\beta}} \right) B(z)~,
\label{Eq3.2.2}
\end{eqnarray}
where in the penultimate step we have used the antisymmetry of the matrix $\Omega$.

\subsection{The Wigner function}

Following the strategy of section 3A, we can use the generalized Weyl-Wigner map to define the noncommutative Wigner function (NCWF). Since the SW map is a linear transformation, the Jacobian is a constant and we may absorb it in the definition of the NCWF to get a normalized distribution (c.f. the remark at the end of section 2):

\vspace{0.3 cm}
\noindent
{\bf Definition 3.8:} Let the system be in a pure or mixed state represented by a density matrix $\hat{\rho} \in \hat{\cal F}$. The NCWF in the noncommutative variables $z=(q,p)$ is defined by:
\begin{equation}
f^{NC}(z) \equiv  {1\over ( \det {\bf \Omega} )^{1/2} (2\pi\hbar)^d} W^{\xi}_z (\hat{\rho})~,
\label{Eq3.3.1}
\end{equation}
where $\det \Omega$ is the determinant of (\ref{Eq3.1.5}).$_{\Box}$

\vspace{0.3 cm}
\noindent
It follows from the definition of $f^{NC}(z)$ that if $\xi = (R,\Pi)$ is the set of Heisenberg variables obtained via the SW map (\ref{Eq3.5}) and we  compute the ordinary Wigner function $f^W (\xi) \equiv \frac{1}{(2\pi\hbar)^d} W_{\xi} (\hat{\rho})$ associated with $\hat{\rho}$, then
$f^{NC}(z)=\left(| \det {\bf \Omega} | \right)^{-\half} f^W \left(\xi (z) \right)$, where $\xi (z)$ is the inverse transformation of the SW map.
Moreover, since the SW map is a linear transformation, the Jacobian $\det {\bf \Omega}^{-1/2}$ is a constant, and it follows from Eq. (\ref{Eq2.29}) that the NCWF is normalized:

\begin{equation}
\int dz \hspace{0.2 cm} f^{NC}(z) = 1.
\label{Eq3.3.2}
\end{equation}
Other properties of the NCWF follow directly from application of the generalized Weyl-Wigner map. For instance, if we apply this map to the von-Neumann equation
\begin{equation}
i \hbar{\partial \hat{\rho}\over\partial t} = \left[\hat H, \hat{\rho} \right],
\label{Eq3.3.3}
\end{equation}
we immediately get the following theorem:

\vspace{0.3 cm}
\noindent
{\bf Theorem 3.9:} The dynamics of the NCWF is dictated by the noncommutative von Neumann-Moyal equation:
\begin{equation}
{\partial f^{NC} \over \partial t}(z,t) = \left[H(z),f^{NC}(z,t)\right]_{\star},
\label{Eq3.3.4}
\end{equation}
where
\begin{equation}
H(z) \equiv W^{\xi}_z (\hat{H})(z).
\label{Eq3.3.5}
\end{equation}
Likewise, suppose that $| \psi_a\rangle, | \psi_b\rangle$ are normalized eigenstates of a certain operator $\hat A ( \hat z) \in \hat{{\cal A}} ({\cal H})$ with eigenvalues $a,b$:
\begin{equation}
\hat A | \psi_a \rangle = a | \psi_a \rangle \hspace{0.5 cm}, \hspace{0.5 cm} \hat A | \psi_b \rangle = b | \psi_b \rangle .
\label{Eq3.3.6}
\end{equation}
If we apply the generalized Weyl-Wigner map to the non-diagonal density matrix element $| \psi_a\rangle\langle \psi_b| \in \hat{{\cal F}}$, we obtain the non-diagonal NCWF:
\begin{equation}
f^{NC}_{ab} (z) = \left( \det {\bf \Omega} \right)^{- 1/2} f^W_{ab} ( \xi (z)),
\label{Eq3.3.7}
\end{equation}
where
\begin{equation}
f^W_{ab} (\xi) \equiv {1\over (\pi \hbar)^d } \int dy \hspace{0.2 cm} e^{-2i p \cdot y / \hbar } \langle R+ y| \psi_a\rangle \langle \psi_b | R-y \rangle.
\label{Eq3.3.8}
\end{equation}
We then get:

\vspace{0.3 cm}
\noindent
{\bf Lemma 3.10:} The non-diagonal NCWF (\ref{Eq3.3.7}) is a solution of the $\star$-genvalue equations:
\begin{equation}
A(z)  \star f^{NC}_{ab} (z) = a f^{NC}_{ab} (z)\hspace{0.5 cm}, \hspace{0.5 cm} f^{NC}_{ab} (z) \star A(z) = b f^{NC}_{ab} (z),
\label{Eq3.3.9}
\end{equation}
where $A(z) \equiv W^{\xi}_z (\hat A)$.

\vspace{0.3 cm}
\noindent
{\bf Proof:} This result can be easily proved, if we apply the generalized Weyl-Wigner map to the equations
\begin{equation}
\hat A | \psi_a\rangle\langle \psi_b| = a | \psi_a\rangle\langle \psi_b| \hspace{0.5 cm}, \hspace{0.5 cm} | \psi_a\rangle\langle \psi_b| \hat A = b | \psi_a\rangle\langle \psi_b|
\label{Eq3.3.10}
\end{equation}
and multiply them by $(2 \pi \hbar)^{- d} ( \det {\bf \Omega} )^{- 1/2}$.$_{\Box}$

\vspace{0.3 cm}
\noindent
{\bf Lemma 3.11:} Let $\left\{ | \psi_a\rangle,  \hspace{0.2 cm} a \in I \right\}$ be a complete orthonormal basis of the Hilbert space. Let $\left\{f^{NC}_{ab} (z) , \hspace{0.2 cm} a , b \in I \right\}$ be the non-diagonal NCWF as in Eqs. (\ref{Eq3.3.7}) and (\ref{Eq3.3.8}). Then the following identities hold for all $a,b,c,d \in I$:
\begin{equation}
f^{NC}_{ab} (z) \star f^{NC}_{cd} (z) = {\delta_{bc}\over(2 \pi \hbar)^d | \det {\bf \Omega} |^{1/2}}  f^{NC}_{ad} (z).
\label{Eq3.3.11}
\end{equation}

\vspace{0.3 cm}
\noindent
{\bf Proof:} The non-diagonal density matrix elements $\left\{ | \psi_a \rangle\langle \psi_b|,~ (a,b \in I) \right\}$ obey the orthogonality conditions:
\begin{equation}
| \psi_a \rangle\langle \psi_b| | \psi_c \rangle\langle \psi_d| = \delta_{bc} | \psi_a \rangle\langle \psi_d|.
\label{Eq3.3.12}
\end{equation}
If we multiply these equations by $( 2 \pi \hbar)^{- 2d} ( \det {\bf \Omega} )^{-1}$ and apply the generalized Weyl-Wigner map, we recover (\ref{Eq3.3.11}).$_{\Box}$

\vspace{0.3 cm}
\noindent
{\bf Lemma 3.12:} If a system is in a pure state, then the corresponding NCWF satisfies:
\begin{equation}
f^{NC} (z) \star f^{NC} (z) = {1\over(2 \pi \hbar)^d ( \det {\bf \Omega} )^{1/2}}  f^{NC} (z).
\label{Eq3.3.13}
\end{equation}

\vspace{0.3 cm}
\noindent
{\bf Proof:} If the system is in a pure state then the corresponding density matrix satisfies  $\hat{\rho}^2 = \hat{\rho}$. Again, we multiply this equation by $( 2 \pi \hbar)^{- 2d} ( \det {\bf \Omega} )^{-1}$ and apply the generalized Weyl-Wigner map.$_{\Box}$

\vspace{0.3 cm}
\noindent
{\bf Lemma 3.13:} A phase space function $f^{NC} (z)$ is a NCWF if and only if there exists a complex-valued phase-space function $g(z)$ such that: 

(i) $\int dz \hspace{0.2 cm} | g(z) |^2 =1$, 

(ii) $f^{NC} (z) = g^* (z) \star g(z)$.

\vspace{0.3 cm}
\noindent
{\bf Proof:} A linear, trace-class operator $\hat{\rho}$ is a density matrix if and only if there exists an operator $\hat a$ such that: (1) $tr (\hat a^{\dagger} \hat a ) =1$, (2) $\hat{\rho} = \hat a^{\dagger} \hat a $ \cite{Dias1}. If we apply the generalized Weyl-Wigner map to the second condition, we get:
\begin{equation}
f^{NC} (z) = \left({a^* (z)\over(2 \pi \hbar)^{d/2} ( \det {\bf \Omega})^{1/4} } \right) \star \left({a (z)\over(2 \pi \hbar)^{d/2} ( \det {\bf \Omega})^{1/4}} \right),
\label{Eq3.3.14}
\end{equation}
where obviously $a(z) = W_{\xi}^z (\hat a)$. If we define $g(z) \equiv (2 \pi \hbar)^{- d/2} ( \det {\bf \Omega})^{- 1/4} a(z)$, we recover condition (ii) in the lemma.
Moreover, from Eqs. (\ref{Eq2.23})-(\ref{Eq2.24}), we have:
\begin{eqnarray}
1 &=& tr (\hat a^{\dagger} \hat a) \nonumber\\
&=& {1\over(2 \pi \hbar)^d ( \det {\bf \Omega})^{1/2}} \int dz \hspace{0.2 cm} a^* (z) \star a (z) \nonumber\\
&=&  \int dz \left({a^* (z)\over(2 \pi \hbar)^{d/2} ( \det {\bf \Omega})^{1/4} }\right) \left({a (z)\over(2 \pi \hbar)^{d/2} ( \det {\bf \Omega})^{1/4} }\right)\nonumber\\
&=& \int dz |g(z)|^2.
\label{Eq3.3.15}
\end{eqnarray}
And we recover condition (i) in the lemma. Conversely, suppose that $f^{NC} (z)$ is such that (i) and (ii) hold. Then, let us define:
\begin{equation}
\hat a \equiv (2 \pi \hbar)^{d/2}  (\det {\bf \Omega})^{1/4} \left(W^{\xi}_z \right)^{-1} (a (z)) \hspace{0.5 cm},\hspace{0.5cm} \hat{\rho} \equiv (2 \pi \hbar)^d  (\det {\bf \Omega})^{1/2} \left(W^{\xi}_z \right)^{-1} (f^{NC} (z)),
\label{Eq3.3.16}
\end{equation}
where $\left(W^{\xi}_z \right)^{-1}$ is the inverse generalized Weyl-Wigner map. Then from (i) and (ii), we conclude that $tr (\hat a^{\dagger} \hat a) =1$ and $\hat{\rho} = \hat a^{\dagger} \hat a$. This means that $f^{NC}$ is the generalized Weyl-Wigner transform of a quantum density matrix, i.e. a NCWF.$_{\Box}$

\vspace{0.3 cm}
\noindent
{\bf Theorem 3.14:} The expectation value an observable $\hat A (\hat z) \in \hat{{\cal A}} ({\cal H})$ in a state $\hat{\rho} \in \hat{{\cal F}}$, such that $\hat A \hat{\rho} \in \hat{{\cal F}}$, is evaluated according to:
\begin{equation}
<\hat A > = \int dz \hspace{0.2 cm} A(z) f^{NC} (z),
\label{Eq3.3.17}
\end{equation}
where $A(z) = W_{\xi}^z (\hat A)$ and $f^{NC} (z)$ is the NCWF associated with $\hat{\rho}$.

\vspace{0.3 cm}
\noindent
{\bf Proof:} From Eqs. (\ref{Eq2.23}) and (\ref{Eq2.24}) we have:
\begin{eqnarray}
<\hat A>  = Tr (\hat A \hat{\rho}) &=& {1\over(2 \pi \hbar)^d} \int {dz\over\sqrt{\det {\bf \Omega} }} \hspace{0.2 cm} A(z) \star W_{\xi}^z (\hat{\rho} )\nonumber\\
&=& \int dz  A(z) \star f^{NC} (z) = \int dz  A(z)  f^{NC} (z)._{\Box}
\label{Eq3.3.18}
\end{eqnarray}

\subsection{Independence of $W^{\xi}_z$ from the particular choice of the Seiberg-Witten map}

It may strike the reader that the NCWF as defined by Eq. (\ref{Eq3.3.1}) seems to be explicitly dependent on the SW map. Clearly this is physically unacceptable. Also, from a mathematical point of view this seems somewhat paradoxical. Indeed, suppose a NCWF $f^{NC} (z)$ is a solution of the noncommutative $\star$-genvalue Eq. (\ref{Eq3.3.9}) or of the noncommutative von Neumann-Moyal Eq. (\ref{Eq3.3.4}). Since neither the $\star$-product (\ref{Eq3.1.6}) nor the Moyal bracket (\ref{Eq3.2.2}) depend upon the SW map, this is contradictory with the fact that the solution $f^{NC} (z)$ has the form (\ref{Eq3.3.1}) which apparently depends on the SW transformation. However, this is only apparently so.
In this section we will prove that the extended Weyl-Wigner map is independent of the particular choice of the SW transform. Hence the entire phase space formulation of NCQM is invariant under a modification of the SW map.

Suppose that $\hat{\xi}$ and $\hat{\xi}'$ are two sets of Heisenberg variables (\ref{Eq3.4}), related by a unitary transformation:
\begin{equation}
\hat{\xi} = \hat U \hat{\xi}' \hat U^{\dagger}\hspace{0.5 cm}, \hspace{0.5 cm} \hat U \hat U^{\dagger} =1~.
\label{Eq3.3.19}
\end{equation}
We may define two Weyl-Wigner maps $W_{\xi}$, $W_{\xi'}$ (c.f. (\ref{Eq2.7}) and (\ref{Eq2.13})) and consequently two generalized Weyl-Wigner maps $W^{\xi}_z$, $W^{\xi'}_z$. Since the two SW maps are linear diffeormorphisms, then the unitary transformation (\ref{Eq3.3.19}) is itself a linear diffeomorphism:
\begin{equation}
\xi = \xi (\xi') \hspace{0.5 cm}, \hspace{0.5 cm} \xi (z) = \xi (\xi' (z))~.
\label{Eq3.3.20}
\end{equation}
The two generalized Weyl-Wigner maps act on a generic element $\hat A$ of the algebra, which may be an operator or a density matrix, as:
\begin{equation}
W^{\xi}_z (\hat A) = A (\xi (z)) \equiv A_1 (z) \hspace{0.5 cm}, \hspace{0.5 cm} W^{\xi'}_z (\hat A) = A' (\xi' (z)) \equiv A_2 (z),
\label{Eq3.3.21}
\end{equation}
where
\begin{equation}
A (\xi) = W_{\xi} (\hat A) \hspace{0.5 cm}, \hspace{0.5 cm} A' (\xi') = W_{\xi'} (\hat A)~.
\label{Eq3.3.22}
\end{equation}
The obvious question is then whether $A_1 (z) = A_2 (z)$? From the unitary transformation (\ref{Eq3.3.19}), (\ref{Eq3.3.20}) we may write
\begin{equation}
\hat A (\hat{\xi}) = \hat A (\hat U \hat{\xi'} \hat U^{\dagger}) =\hat U  \hat A (\hat{\xi'})  \hat U^{\dagger},
\label{Eq3.3.23}
\end{equation}
and thus
\begin{equation}
A' (\xi') \equiv W_{\xi'} (\hat A) = W_{\xi'} \left(\hat U  \hat A (\hat{\xi'})  \hat U^{\dagger} \right) = U \star_{\hbar} A (\xi') \star_{\hbar} U^*,
\label{Eq3.3.24}
\end{equation}
where $U= U(\xi') = W_{\xi'} (\hat U)$ and the product $\star_{\hbar}$ is the Moyal product (\ref{Eq2.8}) with respect to the variables $\xi'$. Since the unitary transformation (\ref{Eq3.3.19}) is linear (cf. (\ref{Eq3.5}) and (\ref{Eq3.3.20})), it acts as a local coordinate transformation in phase space \cite{Dias3}, i.e.:
\begin{equation}
A' (\xi') = A (U \star_{\hbar} \xi' \star_{\hbar} U^*) = A (\xi (\xi'))~.
\label{Eq3.3.25}
\end{equation}
Hence
\begin{equation}
A_2 (z) = A' (\xi' (z)) = A (\xi (\xi' (z))) = A (\xi (z)) = A_1 (z)~,
\label{Eq3.3.26}
\end{equation}
where we used (\ref{Eq3.3.20}), (\ref{Eq3.3.21}) and (\ref{Eq3.3.25}). This proves that the result is the same irrespective of the particular SW map.
Hence $W^{\xi'}_z=W^{\xi}_z$.
If we now go back to our definition of the NCWF, we come to the conclusion, that in order for the NCWF $f^{NC} (z)$ to be independent of the SW map, the associated Wigner function $f^W (\xi)$ must be based on a density matrix $\langle R+y| \hat{\rho} |R-y\rangle$ or a wave function $\psi (R)$ which depend explicitly on the SW map. This is explicity illustrated in the next section for the harmonic oscillator (c.f. Eqs. (\ref{Eq4.18})-(\ref{Eq4.24})).

\section{Comparison with other proposals}

All the results presented in section 3 for the deformation quantization of noncommutative systems are valid for an arbitrary number of dimensions and general noncommutativity, i.e. both for position and momentum. The aim of this section is to compare our general results with known proposals that can be found in the literature. In Refs. \cite{Jing,Rosenbaum1,Muthukumar} the authors consider the two dimensional plane with only spatial noncommutativity. In this case, the extended Heisenberg algebra simplifies considerably:
\begin{equation}
\left[\hat q_i, \hat q_j \right] = i \theta \epsilon_{ij} \hspace{0.2 cm}, \hspace{0.2 cm} \left[\hat q_i, \hat p_j \right] = i \hbar \delta_{ij} \hspace{0.2 cm},\hspace{0.2 cm} \left[\hat p_i, \hat p_j \right] = 0 \hspace{0.2 cm},  \hspace{0.2 cm} i,j= 1,2,
\label{Eq4.1}
\end{equation}
where $\epsilon_{12} = - \epsilon_{21} =1$, $\epsilon_{11}= \epsilon_{22}=0$ and $\theta$ is the only noncommutativity real parameter. A possible SW map for this algebra is given by
\begin{equation}
\hat R_i = \hat q_i + {\theta\over (2 \hbar)} \epsilon_{ij} \hat p_j\hspace {0.2 cm}, \hspace{0.2 cm} \hat \Pi_i = \hat p_i.
\label{Eq4.2}
\end{equation}
The Jacobian is simply given by:
\begin{equation}
\det {\bf \Omega} = 1.
\label{Eq4.3}
\end{equation}
The authors in Refs. \cite{Jing,Muthukumar} proposed the following expression for the NCWF:
\begin{equation}
f^{NC} (q,p) = {1\over(\pi \hbar)^2} \int dy \hspace{0.2 cm} e^{-2ip \cdot y / \hbar} \psi(q+y) \star_{\theta} \overline{\psi (q-y)},
\label{Eq4.4}
\end{equation}
where $\psi (q) \in L^2 \left(\bkR^2 , dq \right)$ and the $\star_{\theta}$ product is given by Eq. (\ref{Eq3.1.7.B}) with $d=2$ and $\theta_{ij} = \theta \epsilon_{ij}$:
\begin{equation}
A(q)*_{\theta} B(q) = A(q) e^{\frac{i\theta}{2}\frac{\buildrel{\leftarrow}\over\partial}{\partial q_i} \epsilon_{ij} \frac{\buildrel{\rightarrow}\over\partial}{\partial q_j}} B(q)~.
\label{Eq4.5}
\end{equation}
The following lemma explains how Eq. (\ref{Eq4.4}) relates to our general formula Eq. (\ref{Eq3.3.1}).

\vspace{0.3 cm}
\noindent
{\bf Lemma 4.1:} If we choose the linear transformation (\ref{Eq4.2}) as our SW map, then we can cast our formula (\ref{Eq3.3.1}) in the form (\ref{Eq4.4}).

\vspace{0.3 cm}
\noindent
{\bf Proof:} Let us compute the $\star_{\theta}$ product using the kernel representation (\ref{Eq3.1.14}). We denote by ${\bf E}$ the matrix with entries $\epsilon_{ij}$. Notice that the $\star_{\theta}$-product looks exactly like the usual $\star_{\hbar}$-product, where $q_1,q_2, \theta, {\bf E}$ play the role of $q,p,\hbar, {\bf J}$, respectively. The kernel form (\ref{Eq3.1.14}) is then directly applicable. It thus follows that
\begin{eqnarray}
\psi (q+y) \star_{\theta} \overline{\psi (q-y)} &=& {1\over(\pi \theta)^2} \int d q' \int d q''  \psi (q'+y) \overline{\psi (q''-y)} e^{[ - \frac{2i}{\theta} (q-q')^T {\bf E} (q''-q)]} \nonumber\\
&=& {1\over(\pi \theta)^2} \int d u \int d v  \psi (u) \overline{\psi (v)} e^{[ - \frac{2i}{\theta} (q-u +y)^T {\bf E} (v+y-q)]}.
\label{Eq4.6}
\end{eqnarray}
Substituting this expression in the Eq. (\ref{Eq4.4}) and integrating over $y$, we obtain:
\begin{eqnarray}
f^{NC} (q,p) &=& {1\over(\pi \hbar)^2} \int d u \int d v  \delta \left(v -2 q - \frac{\theta}{\hbar} {\bf E} p + u \right) \psi (u) \overline{\psi (v)} e^{[ - \frac{2i}{\theta} (q-u)^T {\bf E} (v-q)]} \nonumber\\
&=& {1\over(\pi \hbar)^2} \int d u  \psi (u) \overline{\psi \left(2 q + \frac{\theta}{\hbar} {\bf E} p - u \right)} e^{ - \frac{2i}{\hbar} p \cdot (u-q)},
\label{Eq4.7}
\end{eqnarray}
where we have used the antisymmetry of the matrix $\epsilon_{ij}$. Finally we perform the substitution $u = q + \frac{\theta}{2 \hbar} {\bf E} p + y$ to obtain:
\begin{equation}
f^{NC} (q,p) =  {1\over(\pi \hbar)^2} \int d y \psi \left(q + \frac{\theta}{2 \hbar} {\bf E} p +y \right) \overline{\psi \left(q + \frac{\theta}{2 \hbar} {\bf E} p -y \right)} e^{ - 2 i p \cdot y / \hbar}~.
\label{Eq4.8}
\end{equation}
Notice that from Eq. (\ref{Eq4.2}), we can rewrite this as:
\begin{eqnarray}
f^{NC} (q,p) &=& {1\over(\pi \hbar)^2} \int d y \hspace{0.2 cm}  \psi \left( R(q,p) +y \right) \overline{\psi \left(R(q,p) -y \right)} e^{ - 2 i \Pi (q,p) \cdot y / \hbar} \nonumber\\
&=& f^W \left( R(q,p) , \Pi (q,p) \right),
\label{Eq4.9}
\end{eqnarray}
in agreement with (\ref{Eq3.3.1}). We recall that the Jacobian of the transformation is equal to $1$.$_{\Box}$

\vspace{0.3 cm}
\noindent
In Ref. \cite{Rosenbaum1} the authors considered again the plane with spatial noncommutativity (\ref{Eq4.1}). Since e.g. $\hat q_1$ and $\hat p_2$ commute, they derived the deformation quantization of such systems by resorting to the ket basis $\left\{|q_1, p_2\rangle \right\}$. They thus obtained the following expression for the NCWF:
\begin{eqnarray}
f^{NC}(q,p) & \equiv & {1\over(2\pi\hbar)^2} \int dx_1 \int dy_2 \overline{\phi \left(q_1+\frac{x_1}{2}+\frac{\theta y_2}{2\hbar},p_2-\frac{y_2}{2} \right)} \phi \left(q_1-\frac{x_1}{2} - \frac{\theta y_2}{2 \hbar}, p_2 + \frac{y_2}{2} \right)\times \nonumber\\
& \times & e^{\frac{i}{\hbar} x_1 p_1 + \frac{i}{\hbar} y_2 q_2}.
\label{Eq4.10}
\end{eqnarray}
Here $\phi (q_1, p_2)$ is the $q_1, p_2$ representation of some Hilbert space state $|\phi\rangle$, i.e. $ \phi (q_1, p_2) \equiv  \langle q_1, p_2 | \phi\rangle$. Again it is possible to relate their expression to our general formalism:

\vspace{1 cm}
\noindent
{\bf Lemma 4.2:} If we define $\psi (q_1,q_2) \in L^2 \left( \bkR^2 , dq_1 dq_2 \right)$ by
\begin{eqnarray}
\psi(q_1,q_2) = {1\over\sqrt{2\pi\hbar}} \int dp_2 \hspace{0.2 cm} \phi \left(q_1-\frac{\theta}{2\hbar} p_2, p_2 \right) e^{\frac{i}{\hbar}q_2 p_2},\\
\phi(q_1,p_2) = {1\over\sqrt{2\pi\hbar}} \int dq_2 \hspace{0.2 cm} \psi \left(q_1 + \frac{\theta}{2\hbar} p_2,q_2 \right) e^{-\frac{i}{\hbar}q_2 p_2}
\label{Eq4.11}
\end{eqnarray}
then the Ref. \cite{Rosenbaum2} formula (\ref{Eq4.10}) is equivalent to (\ref{Eq3.3.1}) and (\ref{Eq4.4}).

\vspace{1 cm}
\noindent
{\bf Proof:} Let us substitute the expression (\ref{Eq4.11}) into Eq. (\ref{Eq4.10}) and perform the change of variable $x_1= -2 y_1 - \frac{\theta y_2}{2 \hbar}$. The result is:
\begin{equation}
\begin{array}{c}
f^{NC} (q,p) = {2\over(2 \pi \hbar)^3} \int dy_1 \int d y_2 \int d q_2' \int d q_2'' \hspace{0.2 cm} \overline{\psi \left(q_1 + \frac{\theta}{2 \hbar} p_2 - y_1,q_2' \right)} \psi \left(q_1 + \frac{\theta}{2 \hbar} p_2 + y_1,q_2'' \right) \times \\
\\
\times e^{[\frac{i}{\hbar} \left(q_2' p_2 - \frac{1}{2} q_2' y_2 - q_2'' p_2  - \frac{1}{2} y_2 q_2''-2  p_1 y_1 - \frac{\theta}{2 \hbar} p_1 y_2 + y_2 q_2 \right) ]}.
\end{array}
\label{Eq4.12}
\end{equation}
The integration over $y_2$ yields $4 \pi \hbar \delta \left(- q_2' - q_2 '' - \theta p_1 / \hbar + 2 q_2 \right)$. Finally, if we integrate over $q_2''$ and perform the substitution $q_2' = q_2 - \frac{\theta}{2 \hbar} p_1 - y_2$, to obtain:
\begin{eqnarray}
f^{NC} (q,p) &=& {1\over(\pi \hbar)^2} \int dy_1 \int dy_2 \hspace{0.2 cm} \overline{\psi \left(q_1 + \frac{\theta}{2 \hbar} p_2 - y_1, q_2 - \frac{\theta}{2 \hbar} p_1 - y_2 \right)} \times \nonumber\\
&\times& \psi \left(q_1 + \frac{\theta}{2 \hbar} p_2 + y_1,q_2 - \frac{\theta}{2 \hbar} p_1 + y_2 \right) e^{[ - \frac{2i}{\hbar} \left( p_1 y_1 + p_2 y_2 \right)]},
\label{Eq4.13}
\end{eqnarray}
which has the form (\ref{Eq4.8}).$_{\Box}$

\vspace{0.3 cm} \noindent In Ref. \cite{Rosenbaum2} the following strategy for solving
eigenvalue problems in noncommutative phase space is adopted:

(i) Start from a quantum Hamiltonian $\hat H (\hat z)$ written in terms of the noncommutative variables $\hat z = (\hat q, \hat p)$;

(ii) Perform a SW transformation $\hat z \to \hat z (\hat{\xi} )$ and define a new Hamiltonian in terms of the Heisenberg set $\hat{\xi} = (\hat R, \hat{\Pi} )$: $\hat H^{\sharp} (\hat{\xi} ) \equiv \hat H \left( \hat z (\hat{\xi} ) \right)$;

(iii) Apply the ordinary Weyl-Wigner map $W_{\xi}$ to the eigenvalue equation

\begin{equation}
\hat H^{\sharp} (\hat{\xi}) |\psi_E\rangle = E |\psi_E\rangle,
\label{Eq4.14}
\end{equation}
to obtain
\begin{equation}
H^{\sharp} (\xi) \star_{\hbar} f^W_E (\xi) = E f^W_E (\xi),
\label{Eq4.15}
\end{equation}
where, obviously $ H^{\sharp} (\xi) \equiv W_{\xi} (\hat{H}^{\sharp}(\hat \xi))$ and:
\begin{equation}
f^W_E (\xi) \equiv (2 \pi \hbar)^{-d} W_{\xi} \left( | \psi_E\rangle\langle \psi_E| \right) = {1\over(\pi \hbar)^d} \int \hspace{0.2 cm} e^{-2 i \Pi \cdot y / \hbar} \overline{\psi_E (R-y)} \psi_E (R+y).
\label{Eq4.16}
\end{equation}
Examining the diagram 2, we realize that the authors of Ref. \cite{Rosenbaum2} did not complete the square:
$$
\begin{array}{l c c}
\hat B (\hat z) & ------------- \longrightarrow & \hat B^{\sharp} (\hat{\xi})\\
& \hat{SW}^{-1} &\\
& & \downarrow W_{\xi}\\
& & \\
& & B^{\sharp}  (\xi)\\
& &\\
& \mbox{{\bf Diagram 3}} &
\end{array}
$$
as the inverse SW transformation in phase space was not performed. Hence, their solutions are clearly explicitly dependent upon the particular SW map chosen. This is indeed what happens as can be seen from the specific example of the harmonic oscillator
\begin{equation}
\hat H (\hat q, \hat p) = {\hat p^2 \over 2m} + {1 \over 2} m \omega^2 \hat q^2
\label{Eq4.17}
\end{equation}
on the plane and with spatial and momentum noncommutativity
\begin{equation}
\left[ \hat q_i, \hat q_j \right] = i \theta \epsilon_{ij}\hspace{0.2 cm}, \hspace{0.2 cm} \left[ \hat q_i, \hat p_j \right] = i \hbar \delta_{ij}\hspace{0.2 cm}, \hspace{0.2 cm} \left[ \hat p_i, \hat p_j \right] = i \eta \epsilon_{ij}\hspace{0.2 cm}, \hspace{0.2 cm} i,j=1,2.
\label{Eq4.18}
\end{equation}
The following SW map was used:
\begin{equation}
\hat q_i = \lambda \hat R_i - {\theta\over2 \lambda \hbar} \epsilon_{ij} \hat{\Pi}_j \hspace{0.5 cm},\hspace{0.5 cm} \hat p_i = \mu \hat{\Pi}_i + {\eta\over 2 \mu \hbar} \epsilon_{ij} \hat R_j~,
\label{Eq4.19}
\end{equation}
where the parameters $\lambda, \mu$ are subject to the constraint:
\begin{equation}
{\theta \eta \over 4 \hbar^2} = \lambda \mu ( 1 - \lambda \mu )~.
\label{Eq4.20}
\end{equation}
The corresponding Jacobian reads:
\begin{equation}
{\partial (q,p)\over \partial (R, \Pi)} = (\det {\bf \Omega})^{1/2}=1 - {\theta \eta\over\hbar^2}.
\label{Eq4.21}
\end{equation}
The SW transformation is thus invertible provided $\theta \eta < \hbar^2$, in which case we have:
\begin{eqnarray}
\hat R_i &=& \mu \left(1 - {\theta \eta\over\hbar^2} \right)^{- 1 / 2} \left(\hat q_i + {\theta\over 2 \lambda \mu \hbar} \epsilon_{ij} \hat p_j \right)~,\nonumber\\
\hat{\Pi}_i &=& \lambda \left(1 - {\theta \eta\over\hbar^2} \right)^{-1 / 2} \left(\hat p_i-{\eta\over 2 \lambda \mu \hbar} \epsilon_{ij} \hat q_j \right)~.
\label{Eq4.22}
\end{eqnarray}
In our notation, the resulting stargenfunctions (\ref{Eq4.15}) and (\ref{Eq4.16}) are given by:
\begin{equation}
f_{n_1,n_2}^W (R,\Pi) = e^{-\left(\frac{\alpha}{\beta}R^2 + \frac{\beta}{\alpha}{\Pi}^2\right)} L_{n_1} \left(\frac{2}{\hbar}{\Omega}_{+}\right) L_{n_2}\left(\frac{2}{\hbar}{\Omega}_{-}\right),
\label{Eq4.23}
\end{equation}
where $L_n$ are the Laguerre Polynomials, $n_1, n_2$ are non negative integers and
\begin{eqnarray}
{\alpha}^2 &\equiv& {\lambda^2 m \omega^2\over 2} + {\eta^2\over 8m \mu^2 \hbar^2}~,\nonumber\\
{\beta}^2 &\equiv& {\mu^2\over 2m} + {m \omega^2 \theta^2\over 8 \lambda^2 \hbar^2}~,\nonumber\\
{\Omega}_{\pm} &=& {\alpha\over\beta}R^2 + {\beta\over\alpha}{\Pi}^2 \pm 2R {\bf \epsilon} \Pi~.
\label{Eq4.24}
\end{eqnarray}
One can immediately realize is that, as mentioned above, the stargenfunctions depend
explicitly on the parameters $\lambda, \mu$ of the SW map. It is not difficult to see that
if one uses the inverse SW transformation (\ref{Eq4.22}), the constraint (\ref{Eq4.20})
and multiplies by the normalization $(\det {\bf \Omega})^{-1/2} = \left(1- \theta \eta /
\hbar^2 \right)^{-1}$, the resulting NCWF $f_{n_1,n_2}^{NC} (q,p)$ is independent of
$\lambda, \mu$.

\section{Conclusions}

In this paper we have developed the Weyl-Wigner formalism in the context of a phase space
noncommutative extension of quantum mechanics. Our strategy relies on a covariant formulation of deformation quantization, which is based on an extension of the Weyl-Wigner map. Using this formalism, we constructed the extended $\star$-product, the extended Moyal bracket as well as the noncommutative Wigner function. Moreover, we derived several properties of the extended  Weyl-Wigner map and explicitly proved that albeit apparently dependent upon the particular choice of the SW transformation, it is in fact invariant under different choices of such maps. This in agreement with the fact that neither the extended $\star$-product nor the extended Moyal bracket depend on the SW map. The generality of our results allow us to obtain
the noncommutative Wigner function for a large class of 2-dimensional spaces and compare
our results with other models discussed in the literature.

The main advantage of our approach is that it incorporates and generalizes all the previous proposals in a unified framework. In fact, all the models discussed in the literature are particular cases of our formalism, much as for instance, $d=2$, spatial noncommutativity. However, our proposal for the Wigner function (\ref{Eq3.3.1}) is valid for any arbitrary dimension and for both spatial as well as momentum noncommutativity. The results of Refs. \cite{Jing,Rosenbaum1,Muthukumar} are nevertheless useful. We argued that our formula (\ref{Eq3.3.1}) is apparently dependent on the choice of SW map, but not the physical predictions. The expression  (\ref{Eq4.10}) of Ref. \cite{Rosenbaum1} has the advantage of being explicitly independent of the SW map. This is because the authors there resorted to the $|q_1,p_2\rangle$ representation. Moreover, using this representation, their method can be easily extended to the case where noncommutativity of momenta is included. However, we can anticipate some difficulties with this approach, if we consider higher dimensional $(d \ge 3)$ systems. For instance, in $3$ dimensions, we cannot find $3$ observables among $(\hat q_i, \hat p_j,~ i,j=1, 2,3)$ which are mutually commutative, as was the case of $\hat q_1, \hat p_2$ or $\hat q_2, \hat p_1$ in $2$ dimensions. Probably, one will be forced to construct linear combinations of these.

By contrast, as we have shown in Lemma 4.1, the expression (\ref{Eq4.4}), of Ref. \cite{Jing} corresponds to a specific choice (\ref{Eq4.2}) of the SW map. In a forthcoming paper \cite{Bastos1} we will show that this expression also has its advantages. Indeed, in dimension $d$ with spatial noncommutativity:
\begin{equation}
\left[\hat q_i, \hat q_j \right] = i \theta_{ij}\hspace{0.5 cm}, \hspace{0.5 cm} \left[\hat q_i, \hat p_j \right] = i \hbar \delta_{ij}\hspace{0.5 cm}, \hspace{0.5 cm} \left[\hat p_i, \hat p_j \right] = 0\hspace{0.2 cm}, \hspace{0.2 cm} i,j=1, \cdots, d,
\label{Eq5.1}
\end{equation}
we may easily generalize the SW map (\ref{Eq4.2}):
\begin{equation}
\hat R_i = \hat q_i + \frac{1}{2 \hbar} \theta_{ij} \hat p_j\hspace{0.5 cm}, \hspace{0.5 cm} \hat \Pi_i = \hat p_i\hspace{0.2 cm}, \hspace{0.2 cm} i,j =1 , \cdots, d.
\label{Eq5.2}
\end{equation}
Following the proof of Lemma 4.1, we can show that our expression (\ref{Eq3.3.1}) can be cast in the form:
\begin{eqnarray}
f^{NC} (q,p) = \frac{1}{(\pi \hbar)^d} \int dy & e^{- 2i p \cdot y / \hbar} \psi (q+y) \star_{\theta} \psi^* (q-y),
\label{Eq5.3}
\end{eqnarray}
where the $\star_{\theta}$-product is given by (\ref{Eq3.1.7.B}). And so, the formula of Ref. \cite{Jing} is easily generalized to $d$ dimensions with spatial noncommutativity (\ref{Eq5.1}).

In a forthcoming paper \cite{Bastos1} we will further explore the properties of the noncommutative Wigner functions. In particular, we will show that the set of these functions has a non empty intersection with the set of ordinary Wigner functions, even though it is not a subset of the latter. This will allow us to device criteria for assessing whether a transition from noncommutative to ordinary quantum mechanics has taken place. This researh program was initiated in Ref. \cite{Dias5}, where it was derived the noncommutative version of the Hu-Paz-Zhang equation for a Brownian particle linearly coupled to a bath of oscillators at thermal equilibrium on the plane with spatial noncommutativity.

\subsection*{Acknowledgments}

\vspace{0.3cm}

\noindent The work of CB is supported by Funda\c{c}\~{a}o para a Ci\^{e}ncia e a Tecnologia (FCT)
under the fellowship SFRH/BD/24058/2005. The work of NCD and JNP was partially supported by the
grants POCTI/MAT/45306/2002 and POCTI/0208/2003 of the FCT.

\end{document}